\documentclass[a4paper,11pt]{article}

\usepackage{jheppub}

\usepackage[T1]{fontenc}

\usepackage{slashed}

\usepackage{threeparttable}

\usepackage{caption}

\usepackage{subfigure}

\title{\boldmath Dark matter annihilation into leptons through gravity portals}

\author{Xudong Sun}
\author[1]{and Ben--Zhong Dai\note{Corresponding author.}}

\affiliation{School of Physics and Astronomy, Yunnan University, Kunming, 650091, China}
\affiliation{Key Laboratory of Astroparticle Physics, Yunnan Province, Kunming 650091, China}

\emailAdd{bestsunxudong@126.com}
\emailAdd{bzhdai@ynu.edu.cn}

\abstract{
		Dark matter (DM) constitutes 85\% of the matter in the Universe. However, its specific particle property is still unclear. The fundamentals of DM particles subject to gravitational interaction, and that the lepton excess in cosmic rays may originate from DM particles, inspired us to investigate DM particle properties beyond the standard model. We assume that a leptophilic SU(2) doublet exists in nature as the mediator connecting DM with visible leptons. Since general relativity is not renormalizable at the quantum level, it should be regarded as an effective field theory's leading order term. One species of the next-to-leading-order term should be operators linear to the Ricci scalar and containing scalar fields, such as the Higgs field, scalar DM, or the newly introduced SU(2) scalar doublet. These operators can cause DM annihilation through gravity portals. We analyzed constraints from the cosmic antiproton flux, DM relic abundance, cosmic positron flux, cosmic microwave background, and direct detection experiments. The result shows that there is a vast parameter space that is compatible with current experiments. DM with a mass of electroweak scale is only allowed to annihilate into leptons. We further show that the purely gravitational DM better explains the DArk Matter Particle Explorer cosmic lepton excess. Our work provides a promising mechanism for DM particles to connect with standard model particles.
}

\keywords{Cosmology of Theories beyond the SM, Beyond Standard Model}

\begin{document}
	\maketitle
	\flushbottom
	
	\section{Introduction}

	Many satellites and Earth-based instruments make efforts to search dark matter (DM) particles. In the process of detecting cosmic rays, electron and positron excesses have been detected below 2 TeV (between 100 GeV and several TeV) by the High-Energy Antimatter Telescope~\cite{HEAT}, Advanced Thin Ionization Calorimeter (ATIC)~\cite{ATIC}, Payload for Antimatter Matter Exploration and Light-nuclei Astrophysics (PAMELA)~\cite{PAMELA}, Fermi Large Area Telescope (Fermi-LAT)~\cite{FermiLATe+e-}~\cite{FermiLAT7GeVto2TeV}, and Alpha Magnetic Spectrometer (AMS)-02~\cite{AMSe++e-}.
		Many works have tried to use leptophilic DM to explain cosmic electron and positron flux. A DM model with nonstandard thermal history explained the data from PAMELA, ATIC, and FERMI~\cite{Explainleptons1}. Positron spectral features from supersymmetric DM can also explain the rise in the positron to electron ratio at high energies~\cite{Explainleptons2}. A theory of the weakly interacting massive particle (WIMP) with mass 500--800 GeV was proposed to explain the cosmic ray spectra from ATIC and PAMELA~\cite{Explainleptons3}; also reinforced by signals from the Wilkinson Microwave Anisotropy Probe (WMAP) and Energetic Gamma Ray Experiment Telescope (EGRET). A leptophilic DM model was proposed to explain the PAMELA/ATIC excesses and to explain the DAMA annual modulation signal~\cite{Explainleptons4}. A stable Dirac fermion DM was proposed to explain the positron ratio excess observed by PAMELA with a minimal boost factor~\cite{Explainleptons5}. Rather than focusing on a specific particle physics model, a phenomenological approach is taken to fit the PAMELA data with various two-body annihilation channels~\cite{Explainleptons6}. It was also proposed that cosmic ray spectra from PAMELA, Fermi, and ATIC indicate a 700+ GeV WIMP may exist~\cite{Explainleptons7}. After the AMS collaboration released their data, the origin of the high energy cosmic positrons was revisited. It was found that DM models in which 1--3 TeV of DM particles annihilate into observable particles could be responsible for the observed signal~\cite{Explainleptons8}. Using AMS-02 data, an upper limit cross-section of leptophilic DM particles of less than $10^{-24}~\text{cm}^3/\text{sec}$~\cite{Explainleptons9} was obtained. A dynamical DM decay model was proposed to explain the cosmic-ray positron fraction observed by AMS-02, which also satisfies the constraints from FERMI and Planck~\cite{Explainleptons10}. An extension of the Standard Model supplemented by an electroweak triplet scalar field was also proposed. The model accommodates small neutrino masses by the type-II seesaw mechanism, while an additional singlet scalar field can act as leptophilic DM. This DM candidate can account for the positron fraction and flux data from the AMS-02 experiment~\cite{Explainleptons11}. The signatures of the secluded models of WIMP DM, which could also lead to sizeable excess positron flux (even in the absence of astrophysical boost factors) was studied~\cite{Explainleptons12}. The DArk Matter Particle Explorer (DAMPE) performed the latest exploration of the cosmic electron energy spectrum~\cite{DAMPE}, whose data show a broad excess up to TeV energy and a possible line structure at about 1.4 TeV. This excess of electrons/positrons in cosmic rays could be a signal from DM~\cite{DMexplanation}. Assuming that there exists a DM sub-halo near the solar system, many groups use specific DM models to explain cosmic electron and positron excess~\cite{TwomediatorDM}~\cite{explainDAMPE}~\cite{explainDAMPE2}~\cite{explainDAMPE3}~\cite{explainDAMPE4}~\cite{explainDAMPE5}~\cite{GE201888}~\cite{GE2020115140}.
		The CALorimetric Electron Telescope (CALET) collaboration also reported the cosmic electron and positron spectrum to range from 11 GeV to 4.8 TeV ~\cite{CALET1}~\cite{CALET2}.

	Meanwhile, most evidence for the existence of DM comes from its gravitational interaction, including observations of the galactic rotation curve, large scale structure, gravitational lensing effects in collisions of clusters and galaxies, and the cosmic microwave background (CMB) anisotropy~\cite{yu2}. This inspires us to study DM from the perspective of gravity. Given that most evidence for the existence of DM comes from its gravitational effects; the possibility remains that DM is only subject to gravitational interactions.
	
	Since general relativity is not renormalizable at the quantum level, it is conjected that the Ricci scalar and curvature tensor could exist in various operators. Coupling between the Ricci scalar and other scalar fields could be one type of those operators~\cite{BoundsHiggsGravity}~\cite{EffectiveTheory2}~\cite{EffectiveTheory3}. It is found that the coupling between the Ricci scalar and scalar DM could be a connection between the DM and the visible sector. However, in many scenarios~\cite{Oscar}~\cite{DMdecayTGP}~\cite{SunandDai}~\cite{scalarDMfromIF}, leptons do not dominate the decay or annihilation products of DM. Therefore, we assume that a leptophilic SU(2) scalar doublet exists in nature as a possible mediator connecting DM with the visible leptons. This leptophilic SU(2) scalar doublet could also couple to the Ricci scalar. This work investigates the behavior of scalar DM under the influence of these operators.

		To sum up, this paper emphasizes the following three points:
		\begin{itemize}
			\item Based on the fact that there seems to exists lepton excess in the cosmic ray, we assume that DM contributes to this part of the excess.
			\item Based on the fact that most evidence of DM comes from gravitational interaction, this work assumes that DM is only subject to gravitational interaction.
			\item Given that scalar DM decay through a gravity portal receives pressure from observational constraints~\cite{Oscar}, this work considers scalar DM annihilating through a gravity portal.
		\end{itemize}

		We organize this paper as follows. Section~\ref{TheTheory} introduces the model that DM annihilates through a gravity portal and sets the constraints on parameters for the model. Section~\ref{applicationlabel} describes the application of the model.

	\section{Dark matter annihilation through nonminimal coupling to gravity}\label{TheTheory}
	
	\subsection{The effective field theory treatment of quantum gravity}\label{EffectiveTheory}
	
	Despite the great success of DM research in the $\Lambda$-cold-DM ($\Lambda$CDM) model, general relativity is not the correct theory at the quantum level because it is not renormalizable. Hence, general relativity should be regarded as the leading-order term in an effective-field-theory that describes a more fundamental high-energy theory~\cite{BoundsHiggsGravity}. Any diffeomorphism-invariant quantum theory of gravity can be described at energies below the reduced Planck scale $\kappa^{-1}=2.435\times10^{18}~\text{GeV}$ by an effective-field-theory, whose leading-order terms are: $-R/2\kappa^2
	-c_1 R^2
	-c_2 R_{\mu\nu}R^{\mu\nu}$~\cite{BoundsHiggsGravity}~\cite{EffectiveTheory2}~\cite{EffectiveTheory3}, where $R$ is the Ricci scalar, $R_{\mu\nu}$ is the Ricci tensor, and $c_1$ and $c_2$ are coupling constants. Because physicists found the Higgs boson on the Large Hadron Collider, an additional dimension-four operator, $ -\xi_h (H^\dag H)R,$ should be present as the next to leading-order term in the effective theory, where $H$ represents the Higgs doublet, and $\xi_h$ is the coupling constant~\cite{BoundsHiggsGravity}. It is also the lowest dimension operator and the only dimension-four operator among all the operators that Standard Model fields couple to gravity nonminimally. In the theory of purely gravitational DM, it is also indispensable to obtain the correct relic abundance of DM~\cite{ProbingGDM}. Similarly, other scalar fields can also couple to the Ricci scalar and contribute to the effective theory.

	Based on the excess of electrons in cosmic rays, this work assumes a leptophilic SU(2) scalar doublet $\Phi$ that exists in nature and has the same gauge quantum numbers as Standard Model Higgs doublet. We assume this leptophilic SU(2) scalar doublet couple to leptons directly. Naturally, the dimension-four operator $ -\xi_\eta (H^\dag \Phi + \Phi^\dag H) R$ could be present in the effective theory, where $\xi_\eta$ is a coupling constant. Scalar DM could also couple to gravity nonminimally through the operator $ -\xi_{\phi'} \phi R$ and $ -\xi_{\phi} \phi^2 R$, where $\phi$ represents the scalar singlet DM and $\xi_\phi$ and $\xi_{\phi'}$ are coupling constants. The dimension-three operator $-\phi R$ would lead to DM decay. Observations of cosmic rays constrain the DM decay from gravity portal strongly, so the operator $ -\xi_{\phi'} \phi R$ is unlikely to exist~\cite{Oscar}. Therefore, this paper considers the existence of the dimension-four operator $ -\xi_{\phi} \phi^2 R$. It has also been considered by many other works~\cite{scalarDMfromIF}~\cite{ScaleInvariantScalar}~\cite{inflationEWandDM}~\cite{NscalarinflationEWandDM}. Although $-\xi_\Phi\Phi^2 R$ could also exist where $\xi_\Phi$ is a coupling constant, they are not related to the present work. Higher-dimensional operators are suppressed by $\kappa$ and can be neglected at low energies.

	This work found that after electroweak symmetry breaking, two dimension-four operators, $ -\xi_\eta (H^\dag \Phi + \Phi^\dag H) R$ and $ -\xi_\phi \phi^2 R$, act together, which can lead to efficient DM annihilation.
	
	\subsection{Annihilation of DM through a gravity portal}\label{GiDMa}
	
	The action written in the Jordan Frame is:
	\begin{equation}
	\mathcal{S}^{(\text{JF})}
	\supset
	\int d^4x \sqrt{-g}
	[
	-\frac{R}{2\kappa^2}- \xi_\eta (H^\dag \Phi + h.c.) R
	-\xi_\phi \phi^2 R
	-\xi_h H^\dag H R
	+\mathcal{L}_\Phi
	]
	\label{actionJF}
	\end{equation}
	where $g$ is the determinant of the metric tensor $g_{\mu\nu}$, $\kappa=\sqrt{8\pi G}$ is the inverse (reduced) Planck mass, $h.c.$ is the abbreviation of the Hermitian conjugation, $\mathcal{L}_\Phi=\mathcal{T}_\Phi-m_\Phi^2 \Phi^\dag\Phi$, and $\mathcal{T}_\Phi$ is the kinetic term of the SU(2) doublet $\Phi$.
	
	By performing the Weyl transformation of Eq.~\ref{WeylTrans} on the metric tensor, $\Phi$, $H$, and $\phi$ can be decoupled from the Ricci scalar in the Einstein Frame~\cite{Oscar}~\cite{JFandEF}:
	\begin{equation}
	\tilde{g}_{\mu\nu}=\Omega^2 g_{\mu\nu}
	\label{WeylTrans}
	\end{equation}
	where $\Omega^2=1+2\kappa^2 \xi_\eta (H^\dag \Phi + h.c.)+2\kappa^2 \xi_\phi \phi^2+2\kappa^2 \xi_h H^\dag H$.
	The decoupled action in the Einstein Frame is:
	\begin{equation}
	\mathcal{S}^{(\text{EF})}
	\supset
	\int d^4x \sqrt{-\tilde{g}} [
	-\frac{\tilde{R}}{2\kappa^2}
	+\frac{3}{\kappa^2} \frac{\Omega_{,\rho}\tilde{\Omega}^{,\rho}}{\Omega^2}
	+\tilde{\mathcal{L}}_\Phi]
	\label{actioninEF}
	\end{equation}
	where $\tilde{\mathcal{L}}_\Phi=\Omega^{-2} \tilde{\mathcal{T}}_\Phi
	- \Omega^{-4} m_\Phi^2 \Phi^\dag\Phi$.
	In these expressions, all quantities with a tilde are formed from $\tilde{g}^{\mu\nu}$. The second term of the Lagrangian in Eq.~\ref{actioninEF} contains coupling terms between DM and other scalar fields.
	
	Working in the unitary gauge, after electroweak symmetry breaking, the scalar fields can be expressed as:
	\begin{equation}
	H=
	\frac{1}{\sqrt{2}}
	\begin{pmatrix}
	0 \\
	v+h
	\end{pmatrix}
	,~~~~~~~~~~
	\Phi=
	\begin{pmatrix}
	\omega^+ \\
	\frac{1}{\sqrt{2}} (\eta+iA)
	\end{pmatrix}
	,
	\end{equation}
	$v=246$ GeV is the Higgs field's vacuum expectation value, and $h$ is the Standard Model Higgs boson. The SU(2) scalar doublet $\Phi$ contains two charged scalar fields ($\omega^\pm$) and two neutral fields: the charge conjugation and parity (CP) even scalar $\eta$ and the CP odd scalar $A$. In the unitary gauge, the term $H^\dag \Phi + \Phi^\dag H$ reduces to $\eta(v+h)$, so that although $\Phi$ has four degrees of freedom, only the scalar field $\eta$ couple to the Ricci scalar.
	
	In the Lagrangian in Eq.~\ref{actioninEF}, the first focus should be $ 3\Omega_{,\rho}\tilde{\Omega}^{,\rho}/(\kappa^2\Omega^2)$, because it is greatly enhanced by the Planck mass. After electroweak symmetry breaking, it could be expressed as:
	\begin{eqnarray}
	&&\frac{3}{\kappa^2} \frac{\Omega_{,\rho}\tilde{\Omega}^{,\rho}}{\Omega^2}
	=
	\frac{12 \kappa^2 \xi_\eta \xi_\phi v }{\Omega^4}
	\tilde{g}^{\mu\nu}
	\phi \eta_{,\mu} \phi_{,\nu}
	+
	\frac{3 \kappa^2 \xi_\eta^2 v^2}{\Omega^4}
	\tilde{g}^{\mu\nu}
	\eta_{,\mu}
	\eta_{,\nu}
	+
	\frac{6 \kappa^2 \xi_\eta^2 v}{\Omega^4}
	\tilde{g}^{\mu\nu}
	(
	h
	\eta_{,\mu}
	+
	\eta
	h_{,\mu}
	)\eta_{,\nu}
	\nonumber\\
	&&+
	\frac{3 \kappa^2}{\Omega^4}
	\tilde{g}^{\mu\nu}
	(
	\xi_\eta h
	\eta_{,\mu}
	+
	\xi_\eta\eta
	h_{,\mu}
	+
	2 \xi_\phi \phi
	\phi_{,\mu}
	)
	\times
	(
	\xi_\eta h
	\eta_{,\nu}
	+
	\xi_\eta\eta
	h_{,\nu}
	+
	2 \xi_\phi \phi
	\phi_{,\nu}
	) \nonumber \\
	&&+
	\frac{3 \kappa^2}{\Omega^4}
	\tilde{g}^{\mu\nu}
	(\xi_\eta\eta v+\xi_\eta \eta h+\xi_\phi\phi^2)_{,\mu}
	(2\xi_h v h + \xi_h h^2)_{,\nu}
	+
	\frac{3 \kappa^2}{4\Omega^4}
	\tilde{g}^{\mu\nu}
	(2\xi_h v h + \xi_h h^2)_{,\mu}
	(2\xi_h v h + \xi_h h^2)_{,\nu} \nonumber \\
	\label{intTerm}
	\end{eqnarray}
	The first term on the right-hand side of Eq.~\ref{intTerm} shows that the scalar DM can annihilate through the $s$-channel, $\phi,\phi\to\eta\to\texttt{final--state particles}$. The final--state particles of all these channels depend on the properties of the scalar mediator, $\eta$. The reason for this exclusive focus on $s$-channels will be given in Section~\ref{sectionBWresonance}.
	The second term on the right-hand side of Eq.~\ref{intTerm} indicates that the scalar field $\eta$ should be renormalized via $\eta\to\zeta\eta$, with $\zeta\equiv(1+6\xi_\eta^2 v^2 \kappa^2)^{-1/2}$. In consequence, the mass of $\eta$ is given by $m_\eta=\zeta m_\Phi$, and all of the $\eta$ couplings should be rescaled accordingly~\cite{GravityHiggsWeakBoson}. The second line of the equation indicates that the process $\phi,\phi\to\eta,h$ can exist. The third line of the equation indicates that the $\phi,\phi\to h,h$ channel can exist. It turns out to be the key channel for the scalar DM to acquire its correct relic density~\cite{ProbingGDM}.

	\subsection{The properties of the scalar mediator}\label{annihilationchannels}

	\subsubsection{Implications from cosmic rays}
	As mentioned previously, there is an excess of leptons in the cosmic ray, and the annihilation properties of DM depend on the properties of $\eta$. Hence, one can assume that $\eta$ couple to leptons as follows:
	\begin{equation}
	\mathcal{L}_{\text{direct}}
	=
	\xi_e (\bar{L}_e \Phi e_R + h.c.)
	+
	\xi_\mu (\bar{L}_\mu \Phi \mu_R + h.c.)
	\end{equation}
	We use $L_e$, $L_\mu$, and $L_\tau$ to denote the three generations of SU(2) doublet pairs of leptons and use $e_R$, $\mu_R$, and $\tau_R$ to denote the three generations of right-handed leptons. Coupling constants $\xi_e$ and $\xi_\mu$ should be large enough, so that lifetime of particles excited by $\Phi$ is short enough and does not affect the evolution of the early Universe.
	
	In the Einstein Frame, an action involving properties of $\eta$ becomes:
	\begin{equation}
	\mathcal{S}^{(\text{EF})}
	\supset
	\int d^4x \sqrt{-\tilde{g}}
	[
	\mathcal{L}_{\text{direct}}
	\frac{1}{\Omega^4}
	]
	\label{etainEF}
	\end{equation}
	
	After electroweak symmetry breaking, the $\eta$-related part of the interaction term $\mathcal{L}_{\text{direct}} $ in Eq.~\ref{etainEF} simplifies to:
	\begin{equation}
	\frac{\xi_e\bar{e}\zeta\eta e+\xi_\mu\bar{\mu}\zeta\eta \mu}{\sqrt{2}\Omega^4}
	=
	\frac{\xi_e\bar{e}\zeta\eta e+\xi_\mu\bar{\mu}\zeta\eta \mu}{\sqrt{2}}
	+\mathcal{O}(\kappa^2)
	\label{intTerm2}
	\end{equation}
	where $e=e_R+e_L$, $e_L$ is the left-handed lepton. By reading off from Eqs.~\ref{intTerm} and~\ref{intTerm2}, Feynman rules for  many vertices are given in table~\ref{vertexrules}.
	\begin{table}
		\centering
		\caption{\label{vertexrules} Feynman rules in the Einstein Frame}
		\begin{tabular}{ |l|l|}
			\hline
			physical process
			& Feynman rules \\
			\hline
			$\phi,\phi \rightarrow \eta$
			& $12 i \kappa^2 \xi_\eta \xi_\phi \zeta v
			(p_{\phi1} p_\eta+p_{\phi2} p_\eta)$ \\
			\hline
			$ \eta \to \bar{e},e $
			& $i\xi_e\zeta/\sqrt{2}$\\
			\hline
			$ \eta \to \bar{\mu},\mu $
			& $i\xi_\mu\zeta/\sqrt{2}$\\
			\hline
			$ \phi,\phi \to \phi,\phi $
			& $48i\xi_\phi^2\kappa^2(-p_{\text{in1}}p_{\text{in2}}-p_{\text{out1}}p_{\text{out2}}+p_{\text{in1}}p_{\text{out1}}+p_{\text{in1}}p_{\text{out2}}+p_{\text{in2}}p_{\text{out1}}+p_{\text{in2}}p_{\text{out2}})$\\
			\hline
			$ \eta,h \to \eta,h $
			& $6i\xi_\eta^2\kappa^2(2p_{h\text{[in]}}p_{h\text{[out]}}+2p_{\eta\text{[in]}}p_{\eta\text{[out]}}-p_{h\text{[in]}}p_{\eta\text{[in]}}-p_{h\text{[out]}}p_{\eta\text{[out]}}+p_{h\text{[in]}}p_{\eta\text{[out]}}+p_{h\text{[out]}}p_{\eta\text{[in]}})$\\
			\hline
			$ \phi,\phi \to \eta,h $
			& $12i\xi_\phi\xi_\eta\kappa^2(p_{\phi1}p_\eta+p_{\phi2}p_\eta+p_{\phi1}p_h+p_{\phi2}p_h)$\\
			\hline
		\end{tabular}
		\begin{tablenotes}
			\item In the Feynman rules, the symmetry factors are taken into account.
		\end{tablenotes}
	\end{table}

	\subsubsection{Constraints on the mass of the mediator from interactions in its kinetic term}

		Note that the kinetic term of the Lagrangian $\mathcal{L}_\Phi=\mathcal{T}_\Phi-m_\Phi^2 \Phi^\dag\Phi$ contains interactions between electroweak gauge bosons and particles excited by the SU(2) doublet $\Phi$.

		The couplings of two scalars and one vector boson read
		\begin{eqnarray}
		\xi_{Z\eta A}=-\frac{m_Z}{v}(p^\mu_A-p^\mu_\eta) \nonumber\\
		\xi_{W^\mp\eta \omega^\pm}=\mp\frac{i m_W}{v}(p^\mu_{\omega^+}-p^\mu_\eta) \\
		\xi_{W^\mp A \omega^\pm}=\frac{m_W}{v}(p^\mu_{\omega^+}-p^\mu_A) \nonumber
		\end{eqnarray}
		Specifically, the existence of the process $Z\to \eta, A (A\to l^+,l^-,\eta\to l^+,l^-)$ (with $l$ the leptons) would change the decay branch ratio of the $Z$ boson, which would contradict the accurate measurement of the branching ratio of the decay of the $Z$ boson~\cite{reviewPP}. This point naturally sets constraints on the mass of the particles excited by the SU(2) doublet $\Phi$ that $m_\Phi>m_Z/2$ (implying that $m_\eta>m_Z/2$). Besides, there are no couplings of one scalar and two vector bosons.

	\paragraph{A note on the direct detection experiments}

		Direct detection experiments of DM have almost no constraints on the mass of $\eta$. Direct detection experiments aim to monitor the nuclei recoil or electron recoil to find DM particles. In our model this process could be induced by the process $\phi,N\to \phi,N$ or $\phi,e^-\to \phi,e^-$ through a $t$-channel or $u$-channel mediated by $\eta$, where $N$ denotes a nucleus. Unfortunately, these processes are suppressed by the heavy mediator $\eta$. Scattering a cross-section of these processes would receive a suppression factor of $1/m_\eta^{4}$ where $m_\eta>m_Z/2$. For example, the Beam--Dump eXperiment aims to find DM particles with a mass below 1 GeV~\cite{BDeXperiment}. Also, the heavy photon search experiment at the Jefferson Lab only has sensitivity in the mass range below 1 GeV~\cite{HPSexperiment}~\cite{HPSexperiment2}. As analyzed above, the process $\phi,e^-\to \phi,e^-$ receives a suppression factor $1/m_\eta^4$. The experiment also wants to produce DM particle pairs through the $\eta$-on-shell process $e^-,N\to e^-,\eta$ then $\eta\to \phi,\phi$, or through the $\eta$-off-shell process $e^-,N\to e^-,\eta\to e^-,(\phi,\phi)$ with a mediator $\eta$. The on-shell process is hard to achieve since the mass of $\eta$ is heavier than $m_Z/2$; the energies of the initial electrons are insufficient to support the process. The off-shell process is also negligible since it receives an approximate suppression factor of $1/m_\eta^4$.

	\subsubsection{Constraints on the mediator from the observation of the cosmic antiproton flux}
	
	\begin{figure}[tbp]
		\centering
		\includegraphics[scale=0.6]{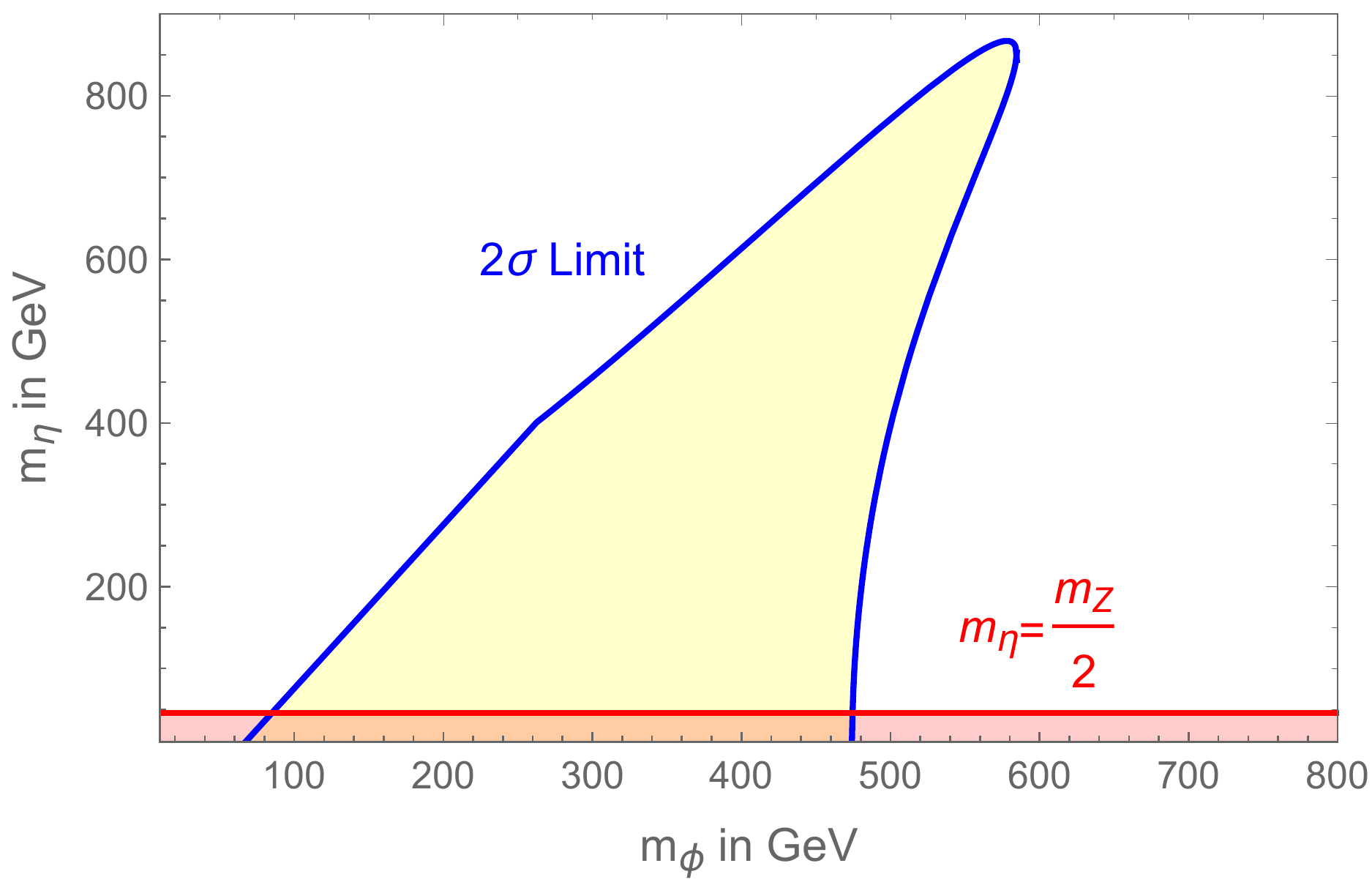}
		\caption{\label{EXantiproton} In the figure, the yellow region surrounded by a blue line is the excluded parameter space by the observation of cosmic antiproton flux when $\langle\sigma v\rangle_{\phi,\phi\to\eta,h}=3\times10^{-26}~\text{cm}^3/\text{s}$. The excluded region $(m_\eta<m_Z/2)$ derived from the kinetic term of $\Phi$ is also shown with the red region for comparison.
		}
	\end{figure}

		Annihilation of DM particles can leave particles in the cosmic rays. Space satellites or Earth-based equipment can detect these signals. Specific to this paper's model, the process $\phi,\phi\to \eta,h$ can leave antiprotons at least through the Higgs boson production. So, the observation of antiprotons can set constraints on DM~\cite{antiprotonlimits1}~\cite{antiprotonlimits2}. The velocity averaged annihilation cross-section should be about $3\times 10^{-26} \text{cm}^3/\text{s}$ to obtain correct relic abundance for the thermally produced DM. Following the results provided by ~\cite{antiprotonlimits3}, we get the excluded two-dimensional parameter space $(m_\phi,m_\eta)$ at a $2\sigma$ confidence level (C.L.) as shown in Fig.~\ref{EXantiproton}. Since we assume $\eta$ tends to decay into leptons, the antiprotons are mainly contributed by the Higgs boson $h$. In the Fig.~\ref{EXantiproton}, we assumed that $\langle\sigma v\rangle_{\phi,\phi\to h,\eta}=3\times10^{-26}~\text{cm}^3/\text{s}$ and the Navarro--Frenk--White (NFW) galactic DM distribution~\cite{NFWprofile}~\cite{NFWprofile2} is used. The excluded region shows that for DM particles with a mass of electroweak scale, the on-shell production of $\eta$ is excluded. This inspires us to consider another possibility, that DM particles connect with standard model particles through the $\eta$-off-shell process, $\phi,\phi\to\eta\to \bar{l},l$, with $l$ the leptons.

	\subsection{The mediator-off-shell process and the DM relic abundance}\label{sectionBWresonance}

	Although TeV DM still allows the on-shell production of $\eta$, this section will consider another special case that $\eta$ is off-shell acting as a mediator connecting DM and leptons. Let us consider the $s$-channel process $\phi,\phi\to\eta\to \bar{l},l$. In the following we assume that $m_\phi\approx m_\eta/2$. Since the DM particle's mass is about half of the mediator $\eta$, namely, $m_\phi\approx m_\eta/2$, the cross-section of DM is enhanced through Breit--Wigner resonance and is sensitive to temperature. Breit--Wigner resonance is the phenomenon that when two particles annihilate through the $s$-channel, if the total energy of the center-of-mass frame is close to the mass of the propagator, the cross-section will be greatly enhanced~\cite{TwomediatorDM}~\cite{BWresonance}. The Breit--Wigner resonance has the potential to solve the problem that the thermally averaged cross-section needed to explain the excess of leptons in the cosmic ray is much larger than the cross-section needed to obtain correct DM relic density~\cite{DMexplanation}. These lepton-production processes could be $\phi,\phi\to\eta\to e^+,e^-$ and $\phi,\phi\to\eta\to \mu^+,\mu^-$. It is also not necessary to worry that the vertex from the kinetic term of $\mathcal{L}_\Phi=\mathcal{T}_\Phi-m_\Phi^2 \Phi^\dag\Phi$ would lead to the production of hadron, since the particles excited by the field $\Phi$ always appear in pairs in a vertex. Consequently, those processes involving the kinetic term would be kinematically forbidden. For example, the process $\phi,\phi\to\eta\to Z ,A$ is kinematically forbidden since $2m_\phi\approx m_\eta\approx m_A\approx m_\Phi$ where $m_A$ is the mass of $A$.

		In the model, the cross-section of the channels $\phi,\phi\to\eta\to e^+,e^-$ and $\phi,\phi\to\eta\to \mu^+,\mu^-$ can be very small at the freeze-out epoch given that the existence of the term $-\xi_h R (H^\dag H)$ supplies an additional gravity portal $\phi,\phi\to h,h$ for DM to annihilate. This portal has been studied in detail in ~\cite{ProbingGDM}. Their research shows that this annihilation channel can reach $3\times10^{-26}~\text{cm}^3/\text{s}$ to obtain the correct DM relic abundance. In the rest of this section, we will show that this annihilation channel also helped our model obtain the correct DM relic abundance.

	\begin{figure}[htbp]
		\subfigure[]{\includegraphics[scale=0.4]{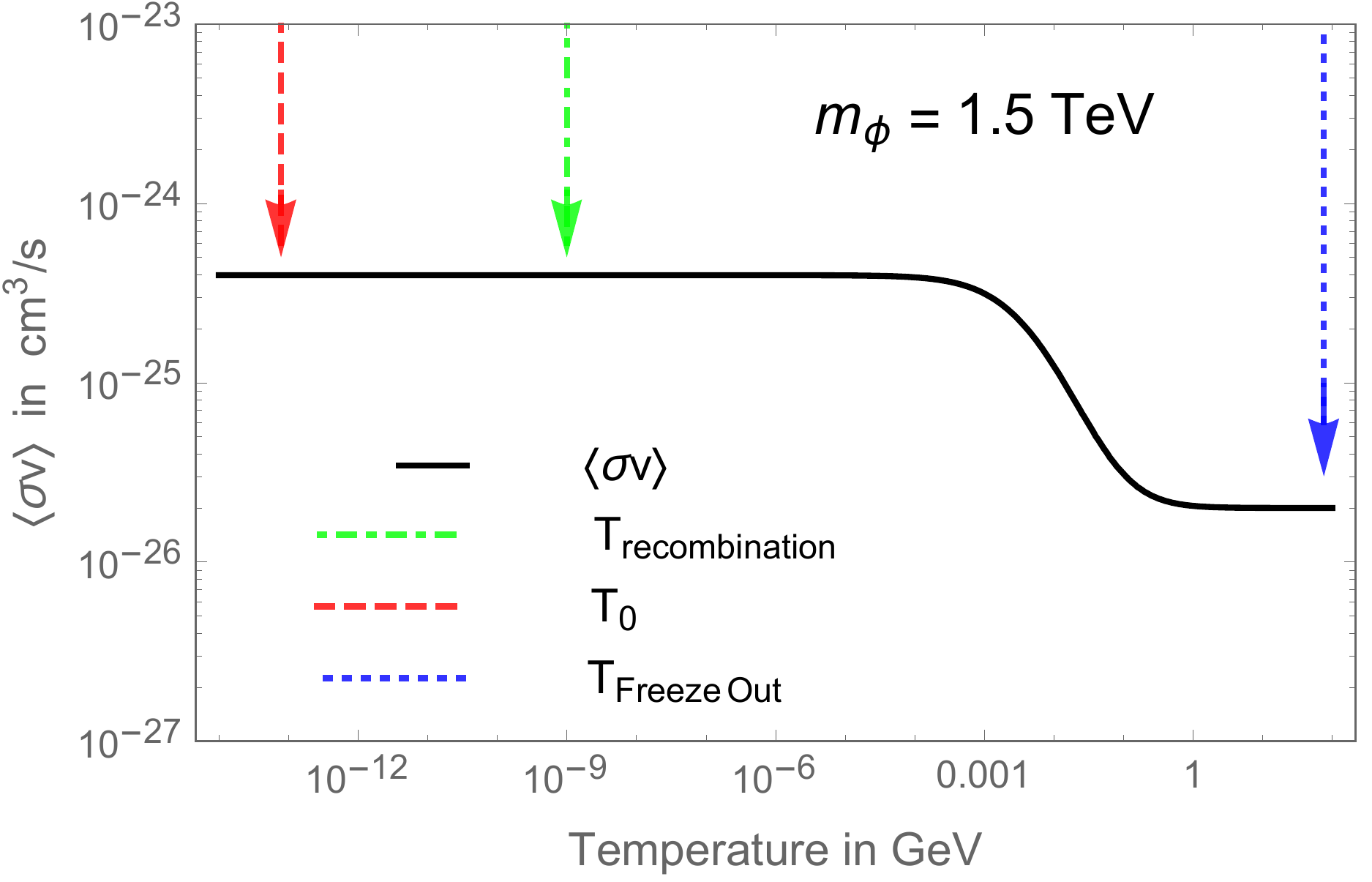}}
		\hfill
		\subfigure[]{\includegraphics[scale=0.4]{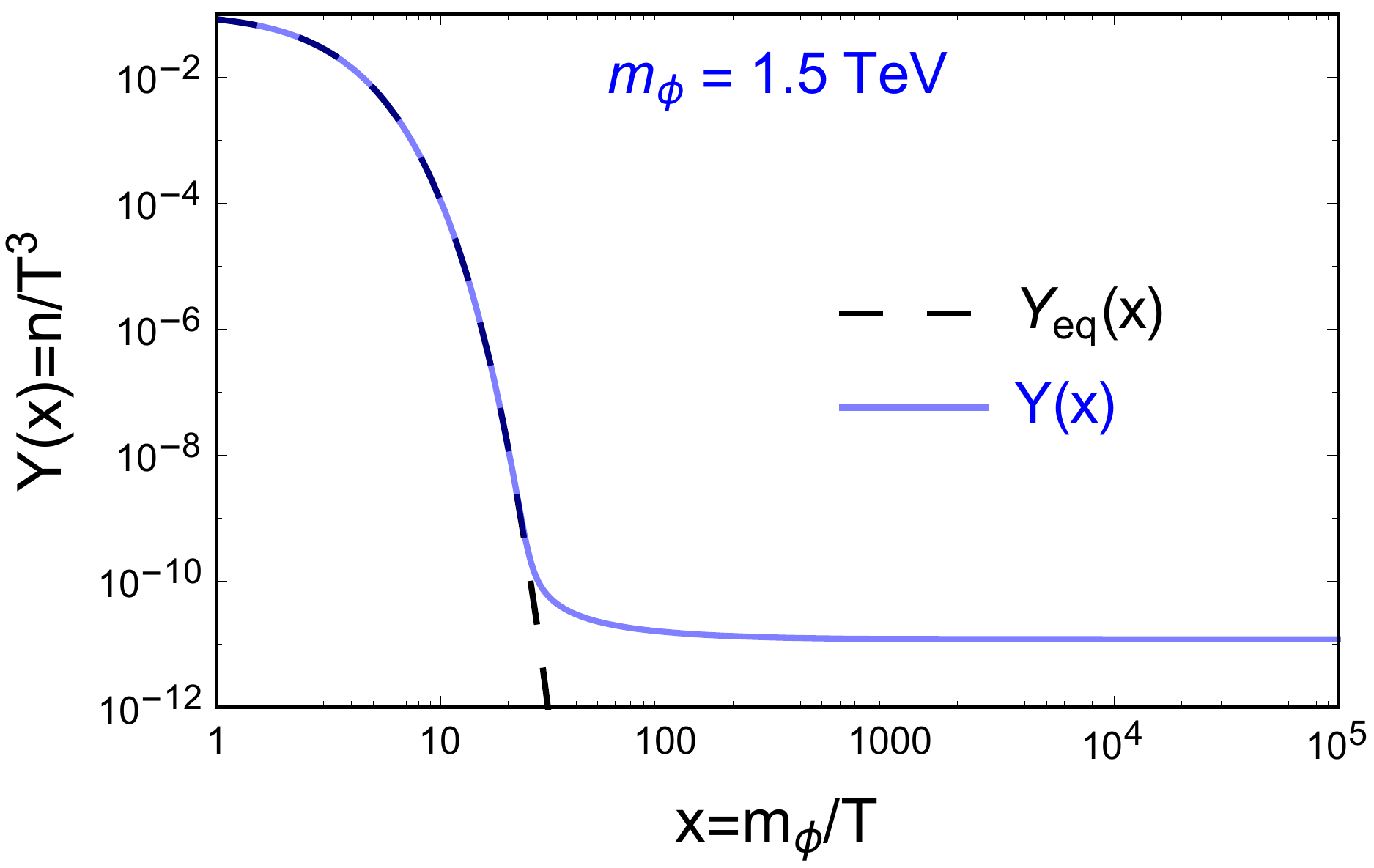}}
		\caption{\label{sensitiveTandRelicAbundance}
			\texttt{(a)} Dependence of total DM thermal average annihilation cross section $\langle\sigma v\rangle$ on DM temperature when $\langle\sigma v\rangle_{\phi,\phi\to h, h}=2\times10^{-26}\text{cm}^3/\text{s}$, $\xi_\eta \xi_\phi \xi_e \zeta^2=2.6\times10^{24}$ and $\xi_\eta \xi_\phi \xi_\mu \zeta^2=1.15\times10^{25}$ (e.g., $\xi_e=2.6\times10^{-4}$, $\xi_\mu=1.15\times10^{-3}$, $\xi_\phi=10^{14}$, $\xi_\eta=10^{14}$), $m_\phi=1500$ GeV, $m_\eta=2m_\phi-\delta$, $\delta=0.01$ GeV. The temperature of the freeze-out epoch has been marked as $T_{\text{Freeze Out}}$, the recombination epoch has been marked as $T_{\text{recombination}}$ and the current DM temperature as $T_0$.\\
		\texttt{(b)} This figure is obtained using the DM cross-section shown in Fig.~\ref{sensitiveTandRelicAbundance} \texttt{(a)}. The solid blue curve shows the comoving DM number density with the evolution of the Universe.
		}
	\end{figure}

		In Fig.~\ref{sensitiveTandRelicAbundance} \texttt{(a)} and \texttt{(b)} we give an example to obtain the correct relic abundance with varying thermal averaged cross-sections. The mass of the propagator $\eta$ is fixed to $m_\eta=2m_\phi-\delta$ with $\delta=0.01~\text{GeV}$ and the mass of DM is fixed to $m_\phi=1500$ GeV. The cross-section of $\phi,\phi\to h, h$ channel is fixed to $\langle\sigma v\rangle_{\phi,\phi\to h, h}=2\times10^{-26}\text{cm}^3/\text{s}$. Fig.~\ref{sensitiveTandRelicAbundance} \texttt{(a)} reveals the enhancement of the velocity averaged cross-section caused by the cooling of the cosmic temperature $T$. In the figure, current $\langle\sigma v\rangle$ of DM is enhanced to $10^{-25}~\text{cm}^3/\text{s}$ to $10^{-24}~\text{cm}^3/\text{s}$. The products from DM annihilation will also play an important role in cosmic rays.
	This is why we focus on $s$-channels in Eq.~\ref{intTerm}.
	
	We will show that by using the cross-section shown in Fig.~\ref{sensitiveTandRelicAbundance} \texttt{(a)}, the correct relic density of DM can be obtained.
		Define $Y(x)=n/T^3$, and $x=m_\phi/T$, where $T$ is the temperature of the Universe, and $n$ is the number density of the scalar singlet. The rate of change of the massive scalar singlet particle density is described by
		\begin{equation}
		\frac{d}{dx}Y(x)=-\frac{\lambda}{x^2}[Y(x)^2-Y_{\text{eq}}(x)^2]
		\label{sigmav1}
		\end{equation}
		where $Y_{\text{eq}}(x)=n^{(0)}/T$ with $n^{(0)}$ the particle number density in thermal equilibrium, and
		\begin{equation}
		\lambda=
		\sqrt{\frac{g_*(m_\phi)}{g_*(T)}}
		\frac{m_\phi^3}{H(m_\phi)} \langle \sigma v \rangle
		\label{sigmav2}
		\end{equation}
		where $H(m_\phi)$ is the Hubble expansion rate when $T=m_\phi$, $\langle \sigma v \rangle $ is the thermal averaged cross-section (it varies with $T$ in our model), and $g_*(T)$ is the total number of degrees of freedom of effectively massless particles (also varies with $T$, and when $T=m_\phi$ it becomes $g_*(m_\phi)$).

		Substituting the cross-section shown in Fig.~\ref{sensitiveTandRelicAbundance} \texttt{(a)} into Eqs.~\ref{sigmav1} and~\ref{sigmav2}, we obtain Fig.~\ref{sensitiveTandRelicAbundance} \texttt{(b)}.
		In Fig.~\ref{sensitiveTandRelicAbundance} \texttt{(b)}, as expected, the number density of the scalar singlet in a comoving volume approaches a constant to $Y_\infty$ when $x$ is large enough. Although total cross-section is enhanced by approximately two orders of magnitude after the freeze-out epoch as shown in Fig.~\ref{sensitiveTandRelicAbundance} \texttt{(a)}, the relic abundance of DM does not change much after the freeze-out epoch ($x\sim23$ for $m_\phi=1500~$GeV) as shown in Fig.~\ref{sensitiveTandRelicAbundance} \texttt{(b)}. Fig.~\ref{sensitiveTandRelicAbundance} \texttt{(b)} reveals that varying $\langle\sigma v\rangle$ can result in the correct DM relic density in our model.

	It is worth explaining the rationale for the large coupling constants $\xi_\eta$ and $\xi_\phi$. Because non-zero $\xi_\eta$ and $\xi_\phi$ do not destroy any symmetries of the Standard Model, the dimensionless coupling constants $\xi_\eta$ and $\xi_\phi$ have no preferred natural values~\cite{BoundsHiggsGravity}~\cite{JFandEF}~\cite{GravityHiggsWeakBoson}. In the Feynman vertices, $\xi_\eta$ and $\xi_\phi$ are always suppressed by the inverse Planck mass. Hence, a large $\xi_\eta$ and $\xi_\phi$ are acceptable as long as they satisfy the perturbation expansion.
	
	\subsection{Bounds on coupling constants}
	\subsubsection{Unitarity bounds}

		This section derives the unitarity bounds on the coupling constants of Eq.~\ref{actionJF}. From the second line of Eq.~\ref{intTerm}, we know that the scattering processes $\phi,\phi\to\phi,\phi$ and $\eta, h\to\eta ,h$ could exist. One can find the corresponding Feynman rules in table~\ref{vertexrules}.

		For the $\phi,\phi\to\phi,\phi$ process, the invariant amplitude is
		\begin{equation}
		\mathcal{M}=96\xi_\phi^2\kappa^2m_\phi^2
		,
		\end{equation}
		for the $\eta, h\to\eta ,h$ process, the invariant amplitude is
		\begin{equation}
		\mathcal{M}=12\xi_\eta^2\kappa^2[m_\eta^2+m_h^2+|\vec{p}_\eta|^2(1-\text{cos }\theta)]
		\label{eq:inv.M}
		\end{equation}
		where $\vec{p}_\eta$ is the three-momentum of $\eta$ in the center-of-mass frame.
		The invariant matrix element in Eq.~\ref{eq:inv.M} depend on the total energy of the center-of-mass frame, which also reflects $- \xi_\eta (H^\dag \Phi + h.c.) R$ is not renormalizable.
		Then we impose the partial wave unitarity condition, $\mathsf{Re}\{a_0\}<1/2$, where
		\begin{equation}
		\mathcal{M}(\theta)=16\pi\sum_{j=0}^{\infty}a_j(2j+1)P_j(\text{cos }\theta)
		\end{equation}
		with $P_j(\text{cos }\theta)$ being the Legendre polynomials that satisfy $P_j(1)=1$.
		We get
		\begin{equation}
		|\xi_\phi|<\frac{\sqrt{\pi}}{2\sqrt{3}}\frac{1}{\kappa m_\phi}
		\end{equation}
		For a typical mass of $m_\phi=2~\text{TeV}$,  $|\xi_\phi|<6.2\times10^{14}$ should hold.
		\begin{equation}
		|\xi_\eta|<\sqrt{\frac{2\pi}{3}}\frac{1}{\kappa \sqrt{m_\eta^2+|\vec{p}_\eta|^2+m_h^2}}
		\end{equation}
		For a typical energy of $m_\eta\approx 2m_\phi=4~\text{TeV}$, $|\xi_\eta|<8.8\times10^{14}$ should hold and $\zeta\equiv(1+6\xi_\eta^2 v^2 \kappa^2)^{-1/2}\approx1$.

		The unitarity bound on $\xi_h$ is also derived as $|\xi_h|<2\sqrt{\pi}/(3\kappa m_\phi)$~\cite{ProbingGDM}. For a typical $m_\phi=2~\text{TeV}$, $|\xi_h|<2.4\times10^{15}$ should hold.

	\subsubsection{Constraints from cosmic positron flux, CMB, and DM relic abundance}

	\begin{figure}[htbp]
		\subfigure[]{\includegraphics[scale=0.4]{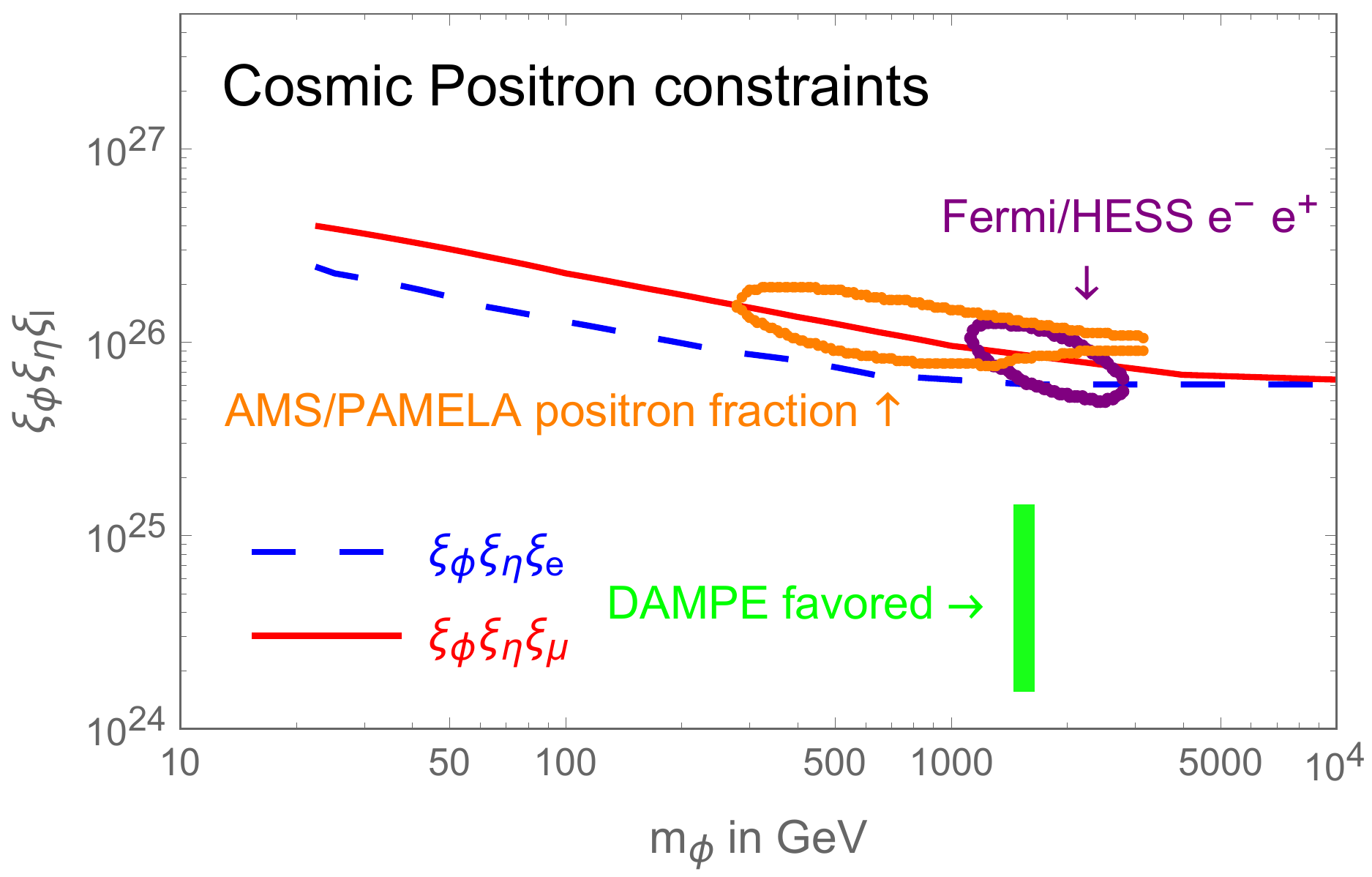}}
		\hfill
		\subfigure[]{\includegraphics[scale=0.4]{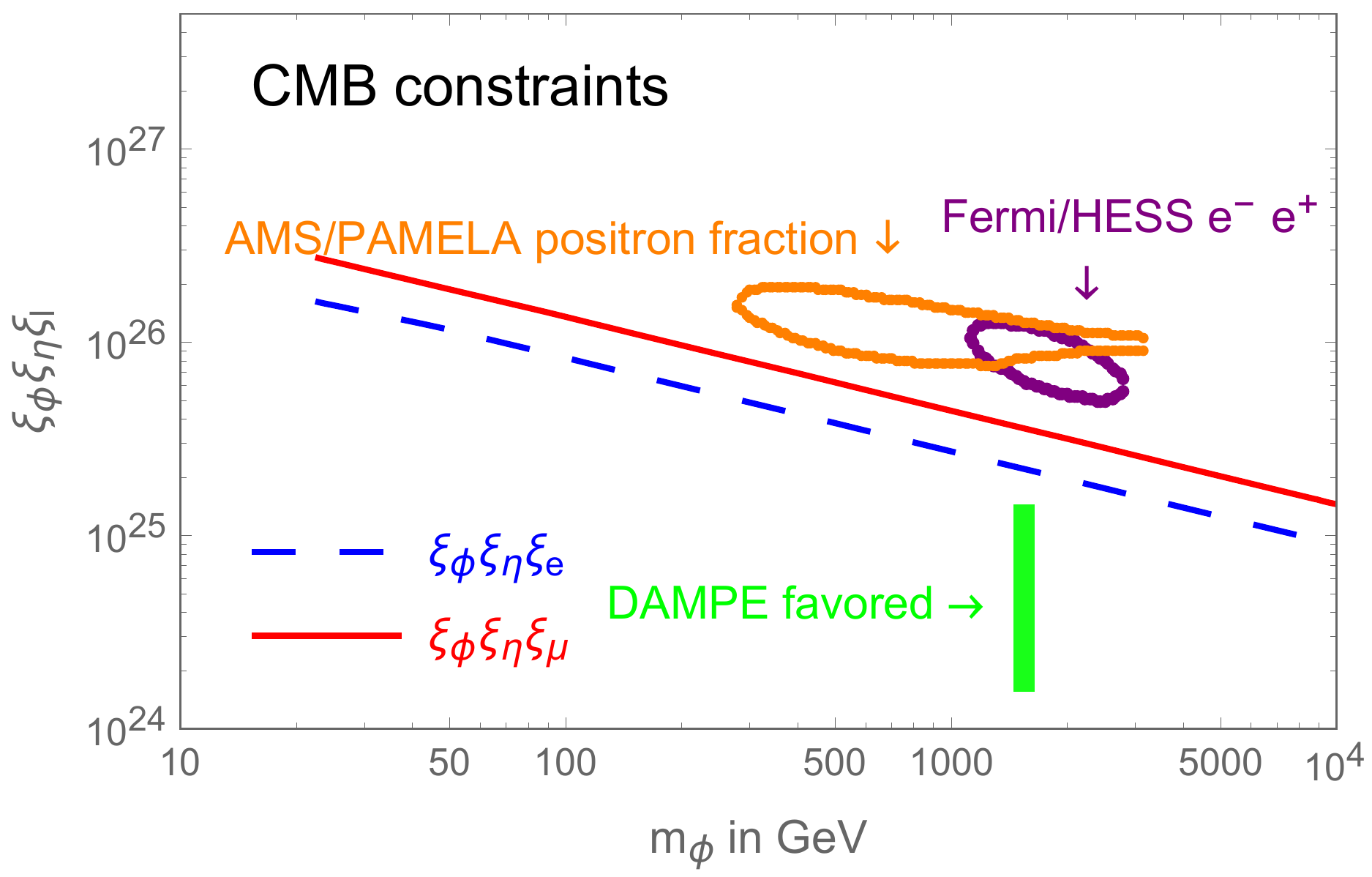}}
		\hfill
		\subfigure[]{\includegraphics[scale=0.4]{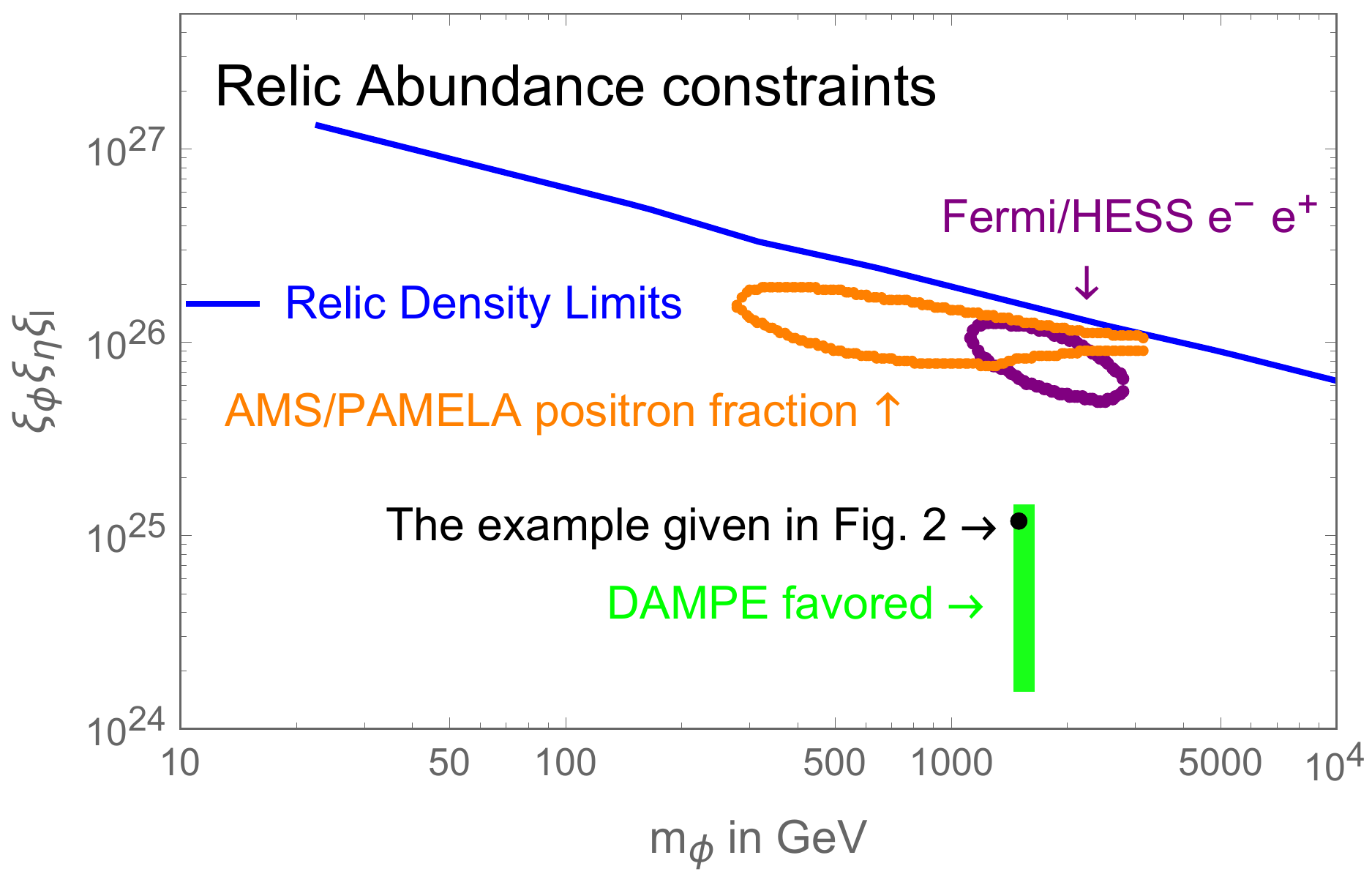}}
		\caption{\label{xiExcluded}
			In all figures, $m_\eta=2m_\phi-\delta$ with $\delta=0.01~\text{GeV}$ are fixed. The green region shows the possible DM explanation of the 1.5 TeV excess reported by DAMPE with $10^{-26}~\text{cm}^3/\text{s}\le\langle\sigma v\rangle_{\phi,\phi\to\eta\to e^+,e^-}\le 10^{-24}~\text{cm}^3/\text{s}$. The $2\sigma$-preferred region suggested by the AMS/PAMELA positron fraction and Fermi/HESS electron and positron fluxes for the leptophilic $\phi,\phi\to\eta\to \mu^+,\mu^-$ channel is surrounded by orange and purple contours, respectively~\cite{Fermi.HESS.AMS.PAMELA}.\\
			\texttt{(a)} The upper limits constrained by the observation of the cosmic positron flux.\\
			\texttt{(b)} The upper limits constrained by the observation of the CMB.\\
			\texttt{(c)} The upper limits constrained by the DM relic abundance.
		}
	\end{figure}

		\paragraph{Positron flux constraints}
		When $m_\eta\approx 2m_\phi$, the $\langle\sigma v\rangle$ is sensitive to the DM temperature, which would result in variation of $\langle\sigma v\rangle$ with the evolution of the Universe. Consequently, current $\langle\sigma v\rangle$ is not determined when $m_\eta\approx 2m_\phi$. Still, the observation of cosmic rays can be used to set limits on the $\langle\sigma v\rangle$, and further set limits on the coupling strength. Specifically, the $\eta$-off-shell process, $\phi,\phi\to\eta\to e^+,e^-$ or $\phi,\phi\to\eta\to \mu^+,\mu^-$, could leave leptons in cosmic rays. The cosmic positron flux is measured by the AMS on the International Space Station~\cite{AMS02}. The predicted positron flux contributed by DM can be calculated via PPPC 4 DM ID~\cite{cookbook}, where we adopted an NFW DM distribution for the Milky Way halo. Provided that the predicted flux cannot exceed the observed positron flux, we could set upper limits on the coupling strength. Define one-sided $\breve{\chi}^2$ as
		\begin{equation}
		\breve{\chi}^2=\sum_i \frac{(F^{\text{th}}_i-F^{\text{obs}}_i)^2}{\delta_i^2}
		\Theta(F^{\text{th}}_i-F^{\text{obs}}_i)
		\label{onesidechi2}
		\end{equation}
		where $F^{\text{th}}_i$ and $F^{\text{obs}}_i$ denote the predicted and observed fluxes, respectively, $\delta_i$ are the experimental errors, and $\Theta(x)$ is the Heaviside function. This work required $\breve{\chi}^2<4$ to obtain an approximate estimate of $2\sigma$ constraint on the coupling strength~\cite{chi2test1}~\cite{chi2test2}.
		Fig.~\ref{xiExcluded} \texttt{(a)} shows the upper limits of $\xi_\phi \xi_\eta \xi_e$ and $\xi_\phi \xi_\eta \xi_\mu$ when $m_\eta=2m_\phi-\delta$ with $\delta=0.01~\text{GeV}$. For comparison, the figure also shows the possible DM explanation of the 1.5 TeV excess reported by DAMPE, with the green region labeled as "DAMPE favored", where the corresponding cross-section is $10^{-26}~\text{cm}^3/\text{s}\le\langle\sigma v\rangle_{\phi,\phi\to\eta\to e^+,e^-}\le 10^{-24}~\text{cm}^3/\text{s}$~\cite{possibleEXofDAMPE}. It can be seen that the "DAMPE favored" region is still alive.

	\paragraph{CMB constraints}

		Annihilation of DM particles would dampen energy into the plasma. Current CMB data are sensitive to energy injection after the recombination epoch, specifically, over redshift between $z\sim600$ and $z\sim1000$ with $z$ as the redshift. The resulting perturbations to the ionization history should be constrained by the CMB temperature and polarization angular power spectra. Using data from the Planck collaboration~\cite{PlanckData2}, Slatyer~\cite{PLANCKconstraints1}~\cite{PLANCKconstraints2} gives a constraint on the velocity averaged cross-section of DM particles, while the Planck collaboration~\cite{PlanckData} gives the most stringent constraints to date. In our model, the Breit--Winger enhanced $\langle\sigma v\rangle$ is a constant after the recombination epoch as shown in Fig.~\ref{sensitiveTandRelicAbundance} \texttt{(a)}. Thermally averaged cross-section of WIMPs annihilation from s-wave channels is also independent of temperature and redshift~\cite{PlanckData}. Therefore, we follow the results provided by the Planck collaboration~\cite{PlanckData} and set constraints on the coupling strength of our model. In Fig.~\ref{sensitiveTandRelicAbundance} \texttt{(b)}, fixing $m_\eta=2m_\phi-\delta$ and $\delta=0.01~\text{GeV}$, we show upper limits of the coupling strength with dashed and solid lines and we show the favored region by the DAMPE resonance at 1.5 TeV with a green box. It can also be seen that the "DAMPE favored" region is alive.

	\paragraph{Relic abundance constraints}

		When DM annihilation benefits from Breit--Winger resonance, namely, the $\langle\sigma v\rangle$ varies with the temperature of the Universe, we are not able to conjecture the specific $\langle\sigma v\rangle$ of DM. Still, the DM relic density can set upper limits on $\langle\sigma v\rangle$ of DM. In Fig.~\ref{xiExcluded} \texttt{(c)} the solid blue line shows the upper limit of $\xi_\phi\xi_\eta\xi_l$ ($\xi_l$ can be regarded as $\xi_l=(\xi_e^2+\xi_\mu^2)^{1/2}$ in the figure) when $m_\eta=2m_\phi-\delta$ with $\delta=0.01~\text{GeV}$. The region above the solid blue line is excluded because larger coupling constants correspond to larger $\langle\sigma v\rangle_{\phi,\phi\to\eta\to e^+,e^-}+\langle\sigma v\rangle_{\phi,\phi\to\eta\to \mu^+,\mu^-}$, which would result in a lack of DM density. The region below the solid blue line is allowed. A small $\xi_\phi\xi_\eta\xi_l$ corresponds to a small $\langle\sigma v\rangle_{\phi,\phi\to\eta\to e^+,e^-}+\langle\sigma v\rangle_{\phi,\phi\to\eta\to \mu^+,\mu^-}$. However, the $\langle\sigma v\rangle_{\phi,\phi\to h,h}$ can be large enough to give a correct relic density~\cite{ProbingGDM}. So, the solid blue line in Fig.~\ref{xiExcluded} \texttt{(c)} corresponds to $\langle\sigma v\rangle_{\phi,\phi\to h,h}=0~\text{cm}^3/\text{s}$. Note that the specific example in Fig.~\ref{sensitiveTandRelicAbundance} corresponds to the black dot in Fig.~\ref{xiExcluded} \texttt{(c)}. We also note that the relic abundance limit is not as stringent as the CMB constraints or cosmic positron constraints. This is because $\langle\sigma v\rangle$ increases with the evolution of the Universe when $m_\eta=2m_\phi-\delta$, $\delta=0.01~\text{GeV}$.

		We conclude that the purely gravitational DM considered here has a wide range of parameter space compatible with current experiments. The purely gravitational DM also has the potential to explain the DAMPE lepton excess.

	\section{Model application}\label{applicationlabel}

	\subsection{Fitting DAMPE with leptons from dark matter}\label{HintsfromDAMPE}
	
	The analysis in Section~\ref{TheTheory} shows that DM could annihilate through a gravity portal and could leave considerable signals in lepton cosmic rays. One of the latest exploration of the cosmic lepton energy spectrum was performed by DAMPE~\cite{DAMPE}; their data show a broad excess in the energy spectrum up to TeV energy and a possible line structure at about 1.4 TeV. Because DAMPE does not discriminate positrons from electrons, we have used lepton to denote electron and positron. As many works proposed, this work also use a DM sub-halo to explain the resonance around 1.4 TeV reported by DAMPE. For a DM halo with an NFW distribution~\cite{NFWprofile}~\cite{NFWprofile2}, the energy density of DM is:
	\begin{equation}
	\rho(r)=\rho_s \frac{(r/r_s)^{-\gamma}}{(1+r/r_s)^{3-\gamma}}
	\label{NFWdistribution}
	\end{equation}
	The following parameters were adopted for the sub-halo: $\gamma=0.5$, $\rho_s=100~\text{GeV}/\text{cm}^3$, and $r_s=0.1~ \text{kpc}$. It was assumed that the distance between the sub-halo center and the solar system was $d_s=0.3~\text{kpc}$. The following parameters were adopted for the Milky Way halo: $\gamma=1$, $\rho_s=0.184~\text{GeV}/\text{cm}^3$, and $r_s=24.42~ \text{kpc}$. It was further assumed that the distance between the Milky Way halo center and the solar system is $r_\odot=8.33~\text{kpc}$.
	
	The mass of the scalar DM was fixed as $m_\phi=1500$ GeV, the mass of the scalar mediator was fixed as $m_\eta=2m_\phi-\delta$, $\delta=0.01$ GeV. Note that three dimensionless coefficients, $\xi_\phi$, $\xi_\eta$, and $\xi_e$, together determine $\langle \sigma v \rangle_{\phi,\phi\to\eta\to e^+,e^-}$, while another set of three dimensionless coefficients, $\xi_\phi$, $\xi_\eta$, and $\xi_\mu$, determines $\langle \sigma v \rangle_{\phi,\phi\to\eta\to\mu^+,\mu^-}$.
	
	The total cosmic electron flux includes three major contributions:
	\begin{equation}
	F^{\text{total}}=F^{\text{BG}}+F^{\text{SH}}+F^{\text{MW}}
	\end{equation}
	where $F^{\text{BG}}$ is the cosmic ray background, $F^{\text{SH}}$ is the contribution of the DM sub-halo, and $F^{\text{MW}}$ is the contribution of the Milky Way DM halo.
		For the cosmic lepton background, we use a single power-law model,
		\begin{equation}
		F^{\text{BG}}=C(\frac{E}{100~\text{GeV}})^{-\alpha}
		.
		\label{PowerLawBG}
		\end{equation}

		Let us consider the contribution from DM annihilation. When we calculate the electron spectrum from the $\phi,\phi\to\eta\to\mu,\mu$ channel, we adopt the approximation that the decay of muon through process $\mu\to e, \bar{\nu}_e, \nu_\mu$ with $100\%$ branching fraction, where $\bar{\nu}_e$ and $\nu_\mu$ denote neutrinos. This paper does not consider the contribution from the $\phi,\phi\to\eta\to\tau^+,\tau^-$ channel. The reason is as follows. $\tau^\pm$ leptons decay through channel $\tau\to e, \bar{\nu}_e, \nu_\tau$ with a $17.83\%$ branching fraction. This component contributes to the same electron spectrum as the decay of muon. $\tau^\pm$ leptons decay through channel $\tau\to \mu, \bar{\nu}_\mu, \nu_\tau $ with a $17.4\%$ branching fraction, and the decay product $\mu^\pm$ decays into electrons and positrons. The contribution of this component to the electron spectrum would appear in the low-energy region. Its contribution to the $E^3 F^{\text{total}}$ will be highly suppressed. Given that the branching fraction of the $\tau\to e, \bar{\nu}_e, \nu_\tau$ channel is only $17.83\%$, it needs many more annihilation events of DM particles for the $\phi,\phi\to\eta\to\tau^+,\tau^-$ channel to contribute the same cosmic electron flux in the high energy region as the $\phi,\phi\to\eta\to\mu^+,\mu^-$ channel. Consequently, in the case of contribute the same lepton flux in the high energy region, DM particles annihilated through the $\phi,\phi\to\eta\to\tau^+,\tau^-$ channel receive more pressure from the observation of cosmic gamma rays than through $\phi,\phi\to\eta\to\mu^+,\mu^-$. Therefore, we are not considering the $\phi,\phi\to\eta\to\tau^+,\tau^-$ channel in this paper.

	Green's function method~\cite{TwomediatorDM}~\cite{GreenFunctionMethod}, was used to calculate the contribution of the DM sub-halo to the cosmic electron flux $F^{\text{SH}}$ and the contribution of the DM Milky Way halo to the cosmic electron flux $F^{\text{MW}}$:
	\begin{subequations}\label{notdecide}
		\begin{equation}
		F^{\text{SH}} (\vec{x},E)+F^{\text{MW}} (\vec{x},E)=\frac{v_e}{4\pi}\int d^3 x_s \int dE_s
		G(\vec{x}, E; \vec{x}_s, E_s)
		Q(\vec{x}_s, E_s)
		\end{equation}
		\begin{equation}
		Q(\vec{x}_s, E_s)=\frac{1}{4}\frac{\rho_\phi^2(\vec{x}_s)}{m_\phi^2} \langle \sigma v\rangle \frac{dN}{dE_s}(E_s)
		\end{equation}
		\begin{equation}
		G(\vec{x}, E; \vec{x}_s, E_s)=\frac{1}{b(E)} (\pi \lambda^2)^{-3/2} e^{-\frac{(\vec{x}-\vec{x}_s)^2}{\lambda^2}}
		\end{equation}
		\begin{equation}
		\lambda^2=4\int_{E}^{E_s}dE'D(E')/b(E')
		\end{equation}
	\end{subequations}
	where the subscript $s$ indicates the quantities associated with the DM source and $b(E)=b_0(E/\text{GeV})^2$ is the energy loss coefficient. The main energy losses are from synchrotron radiation and inverse Compton scattering at energies $E>10$ GeV, where the following parameters are adopted: $b_0=10^{-16}~\text{GeV/s}$, $D(E)=D_0(E/\text{GeV})^\delta$ is the diffusion coefficient, $D_0=11~\text{pc}^2/\text{kyr}$ and $\delta=0.7$, $v_e$ is the velocity of the electrons, $\rho_\phi(\vec{x}_s)$ is the DM mass density, and $dN/dE_s$ is the energy spectrum of electrons per DM annihilation. It can be checked through the above procedure that the contribution of the Milky Way halo to the cosmic electron flux between 500 GeV and 1500 GeV is two orders of magnitude smaller than the contribution of the DM sub-halo.
	
	\begin{figure}[htbp]
		\subfigure[]{\includegraphics[scale=0.43]{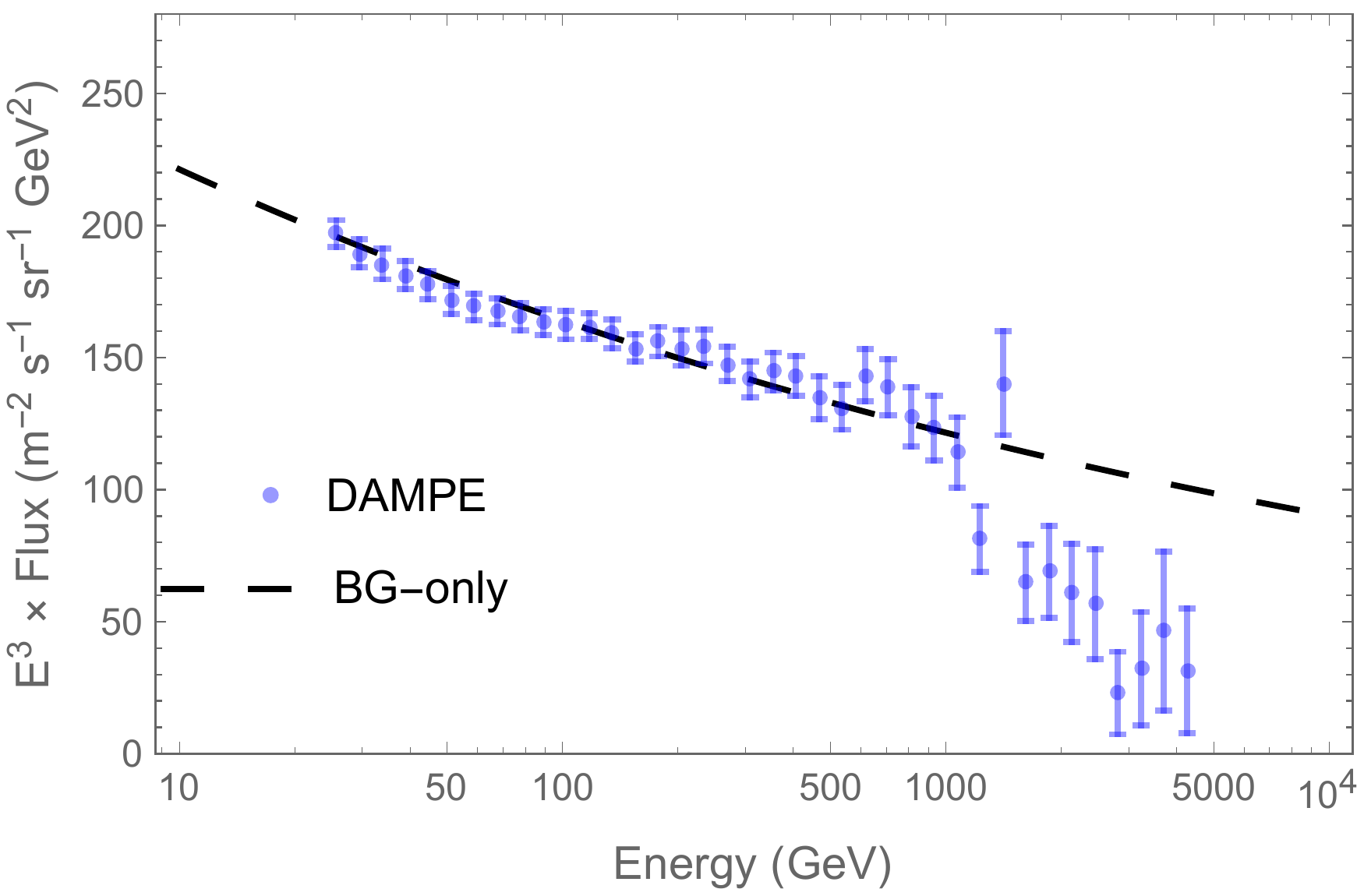}}
		\hfill
		\subfigure[]{\includegraphics[scale=0.43]{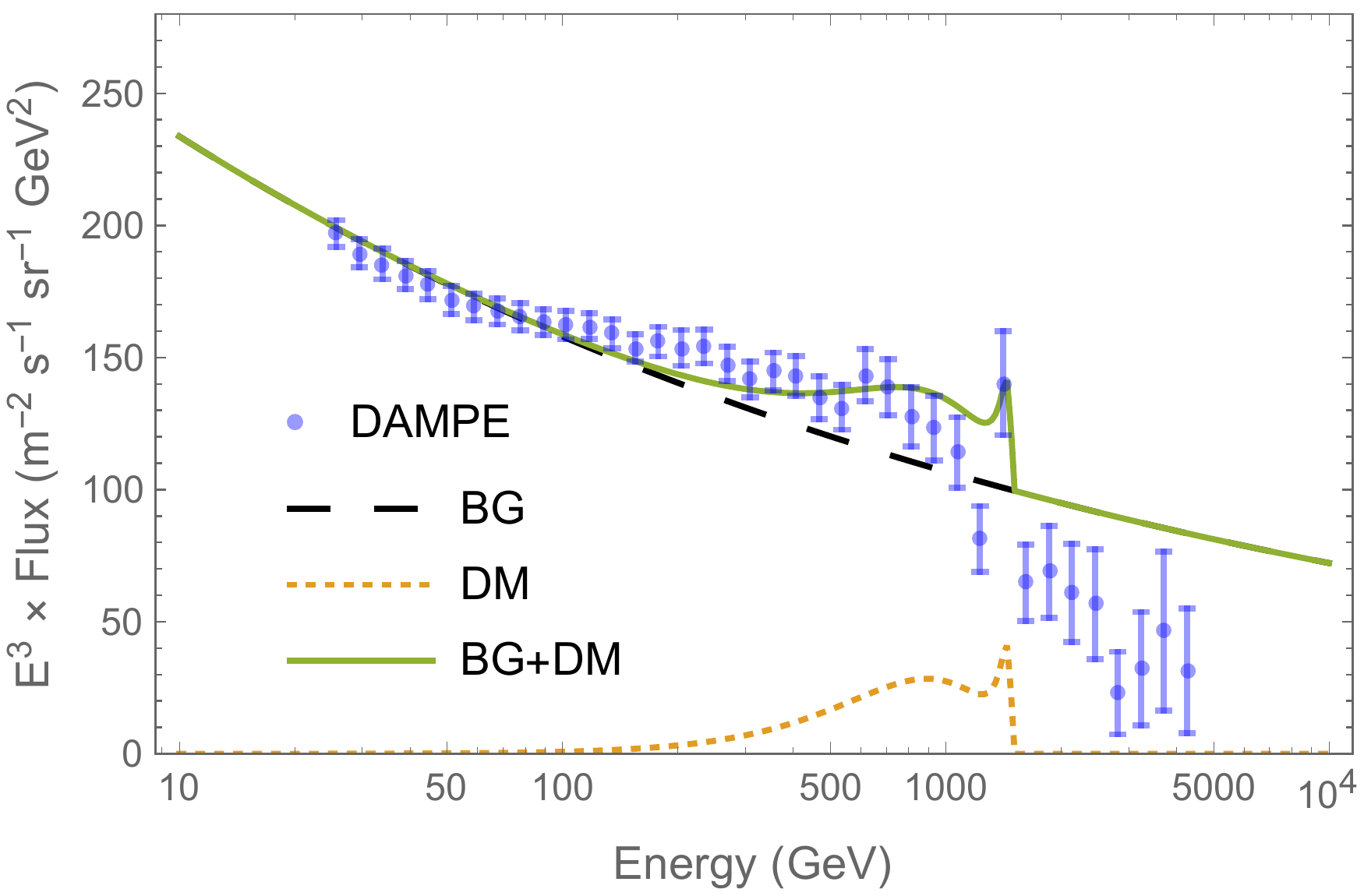}}
		\caption{\label{DAMPEfitandDAMPEfit2}
			Cosmic ray electron spectrum (multiplied by $E^3$). The blue points with error bars are the flux measured by DAMPE.\\
		\texttt{(a)} The dashed line shows a best-fitting power-law spectrum. The fitting quality is $\chi^2/\text{d.o.f.}=101.6/39$.\\
			\texttt{(b)} The solid line shows a combined best-fitting cosmic electron flux. The dashed line shows the cosmic electron background (BG) of the best-fitting. The dotted line shows the flux contributed by DM. The fitting parameters are, $F^{\text{BG}}=C[ E/(100~\text{GeV})]^{-\alpha}$ with $C=1.58\times10^{-4}~(\text{GeV}~\text{m}^2~\text{s sr})^{-1}$ and $\alpha=3.17$, $\langle\sigma v\rangle_{\phi,\phi\to \eta\to e^+,e^-}=1.8\times10^{-26}~\text{cm}^3/\text{s}$, $\langle\sigma v\rangle_{\phi,\phi\to \eta\to \mu^+,\mu^-}=3.6\times10^{-25}~\text{cm}^3/\text{s}$. The fitting quality is $\chi^2/\text{d.o.f.}=82.8/41$.
		}
	\end{figure}

		We use $\chi^2$ analysis to find the best fitting parameters:
		\begin{equation}
		\chi^2=\sum_i \frac{[F_i^{\text{total}}(\langle E\rangle_i^{\text{exp}})-F_i^{\text{exp}}]^2}{\delta_i^2}
		\end{equation}
		where $F_i^{\text{exp}}(\delta_i)$ is the electron flux (uncertainty) of the $i$-th bin as reported by DAMPE or CALET, and $\langle E\rangle_i^{\text{exp}}$ is the representative value of the energy in the $i$-th bin.
		The $\chi^2$ test was applied to the following two cases to fit the DAMPE data:
		\begin{itemize}
			\item [\texttt{(a)}] The background-only case. Lepton flux is contributed by the power-law background, without the contribution from DM annihilation.
			\item [\texttt{(b)}] Combined case. There exists lepton signals from the DM annihilation, with DM mass fixed to $m_\phi=1.5~\text{TeV}$.
		\end{itemize}
		For the background-only case, we use all the points of the DAMPE data to run a $\chi^2$ fit where $F^{\text{BG}}$ follows the form of Eq.~\ref{PowerLawBG}. We obtain the following best-fitting parameters; $C=1.64\times10^{-4}~(\text{GeV}~\text{m}^2~\text{s sr})^{-1}$ and $\alpha=3.13$, where $E$ represents the energy of the electrons. The fitted result is shown in Fig.~\ref{DAMPEfitandDAMPEfit2} \texttt{(a)}, the fitting quality is $\chi^2/\text{d.o.f.}=101.6/39$.

		For the combined case, we do a combined fit (fitting $F^{\text{BG}}$ and $F^{\text{SH}}+F^{\text{MW}}$ simultaneously) to all points in the DAMPE data, with $F^{\text{BG}}$ following the form of Eq.~\ref{PowerLawBG}. The mass of DM is fixed to $m_\phi=1.5~\text{TeV}$. The result is shown in Fig.~\ref{DAMPEfitandDAMPEfit2} \texttt{(b)}. The best fitting parameters are, $F^{\text{BG}}=C[ E/(100~\text{GeV})]^{-\alpha}$ with $C=1.58\times10^{-4}~(\text{GeV}~\text{m}^2~\text{s sr})^{-1}$ and $\alpha=3.17$, $\langle\sigma v\rangle_{\phi,\phi\to \eta\to e^+,e^-}=1.8\times10^{-26}~\text{cm}^3/\text{s}$, which contributes mainly to the resonance at 1.4 TeV, and $\langle\sigma v\rangle_{\phi,\phi\to \eta\to \mu^+,\mu^-}=3.6\times10^{-25}~\text{cm}^3/\text{s}$, which contributes to the broad excess around 500--1000 GeV. The fitting quality is $\chi^2/\text{d.o.f.}=82.8/41$ as shown in Fig.~\ref{DAMPEfitandDAMPEfit2} \texttt{(b)}. The improvement of the fitting quality supports the existence of the DM sub-halo. In order to match the two cross-sections obtained above, we can choose the following parameters: $\xi_\phi=\xi_\eta=10^{14}$; $\xi_e=2.6\times10^{-4}$; $\xi_\mu=1.15\times10^{-3}$; $m_\eta=2 m_\phi-\delta$ with $\delta=0.01~\text{GeV}$. The above parameters show that our model is feasible. Other parameters can also be viable to obtain the same cross-section.

	\subsection{Position of the DM sub-halo (constraints from the IGRB)}\label{IGRBconstraints}

	\begin{figure}[htbp]
		\subfigure[]{\includegraphics[scale=0.27]{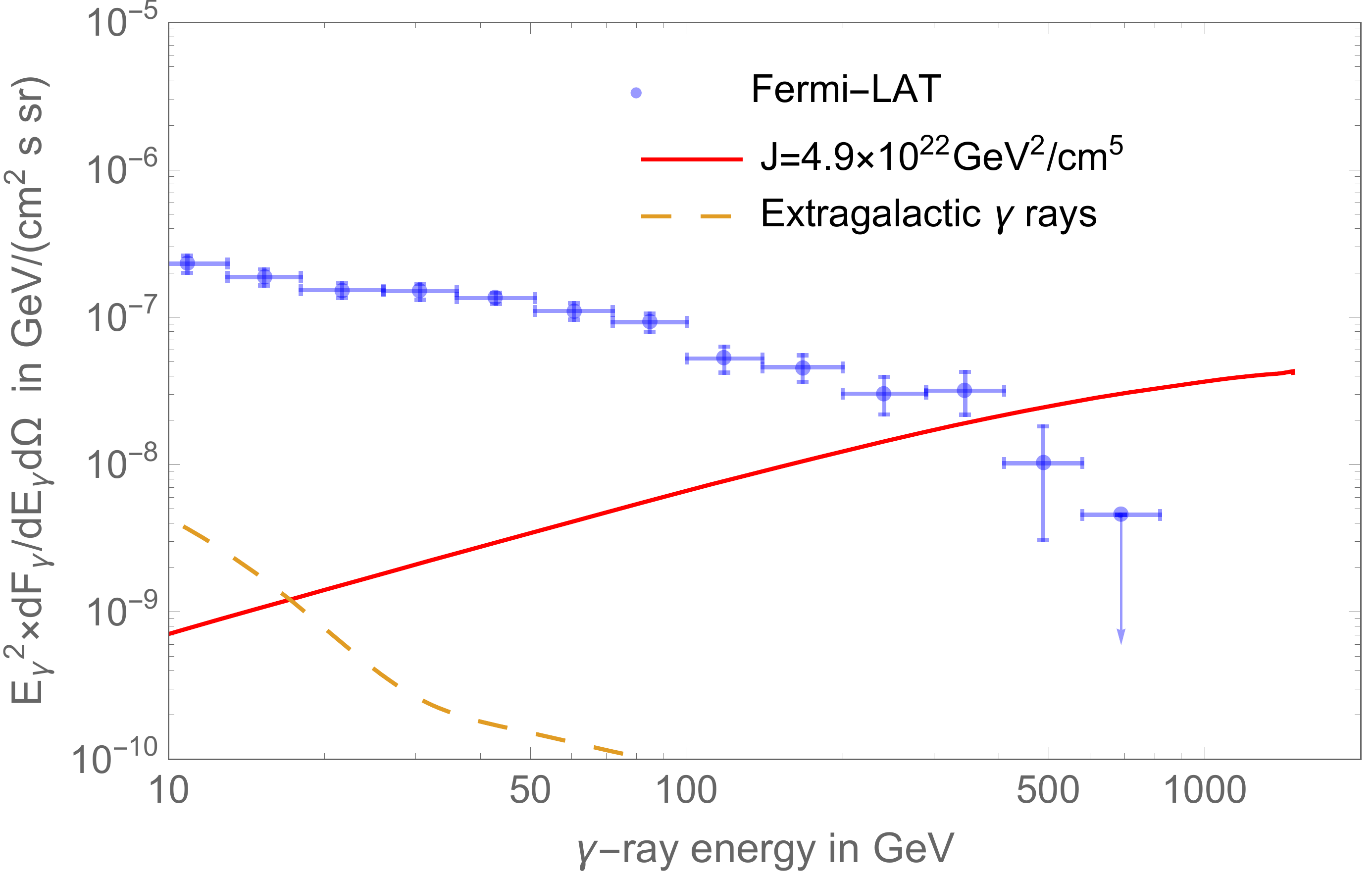}}
		\hfill
		\subfigure[]{\includegraphics[scale=0.40]{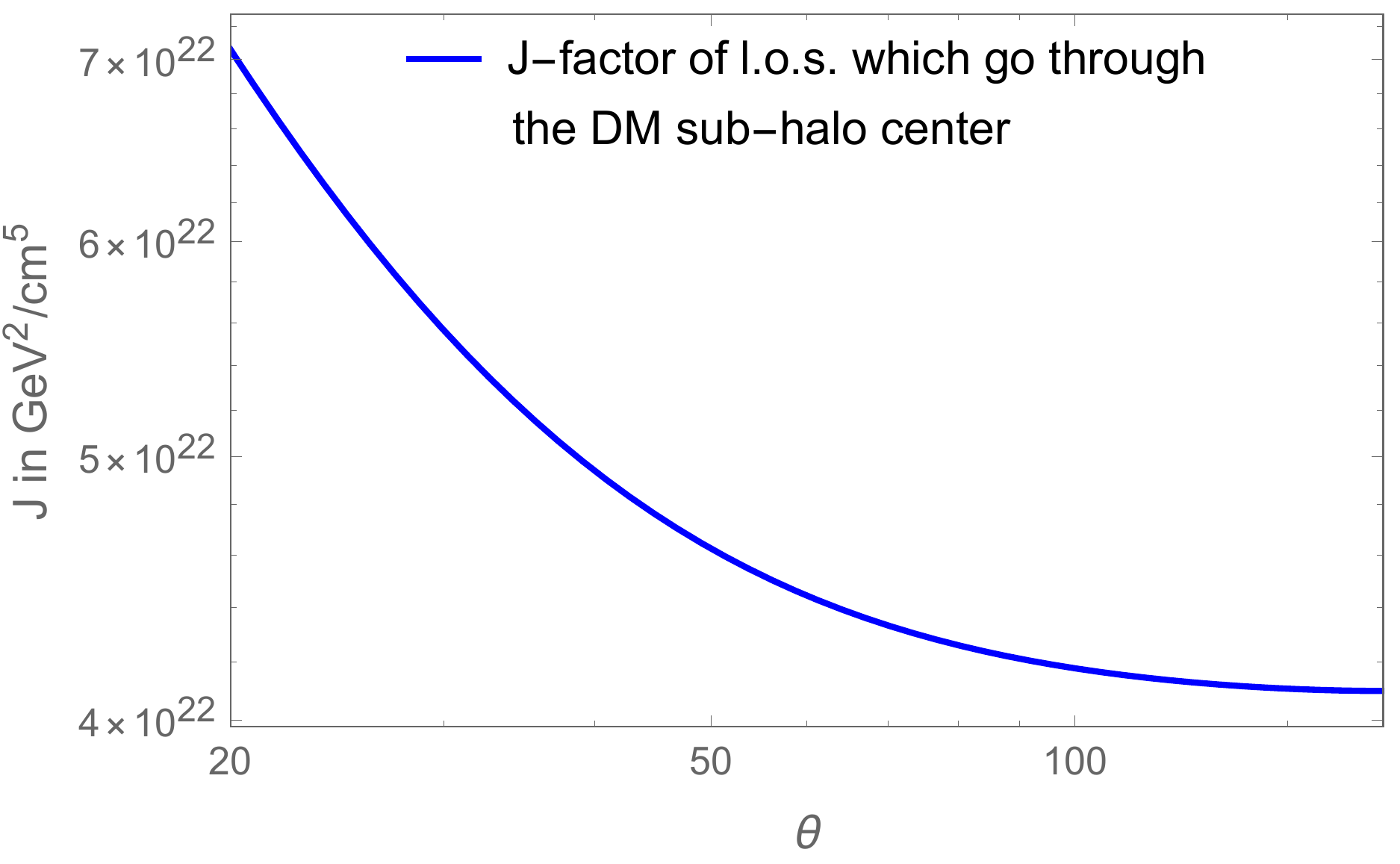}}
		\caption{\label{FermiLATIGRBandJfactor}
			\texttt{(a)} IGRB (multiplied by $E^2$). The blue points with error bars are the flux measured by Fermi-LAT. The solid line shows the prompt emission when the J-factor is $4.9\times10^{22}~\text{GeV}^2/\text{cm}^5$. The dashed line shows the photon flux contributed by extragalactic DM.\\
			\texttt{(b)} In the figure, we fix l.o.s. to go through the DM sub-halo center. So the J-factor changes with the position of the DM sub-halo. In other words, J-factor changes with $\theta$ where $\theta$ is the aperture angle between the direction of the l.o.s. and the axis connecting the Earth to the galactic center.
		}
	\end{figure}
	
	Fermi-LAT measured the IGRB~\cite{FermiLAT}. The $\gamma$-ray flux produced by DM should not exceed the IGRB. The analysis of IGRB by Fermi-LAT included only high latitudes ($|b|>20^\circ$)~\cite{FermiLAT}, where $b$ is the galactic latitude, and $l$ is the galactic longitude.
	The IGRB can set constraints on TeV DM for many windows on the sky~\cite{IGRBonWindow1}~\cite{IGRBonWindow2}~\cite{IGRBonWindow3}. Fermi-LAT could also detect signals from a DM sub-halo in some optimistic cases~\cite{FermiLATsubhalo}.
	
	The prompt $\gamma$-ray flux that can be detected by Fermi-LAT as originating from DM can be calculated as follows:
	\begin{equation}
	\frac{dF_\gamma}{dE_\gamma}
	=
	\frac{\langle\sigma v\rangle}{8\pi m_\phi^2} (\frac{dN_\gamma}{dE_\gamma}) J
	\label{promptgammaray}
	\end{equation}
	where $dN_\gamma/dE_\gamma$ is the energy spectrum of photons per DM annihilation,
		$J$ is the J-factor from a specific line of sight (l.o.s.).
		The J-factor of a l.o.s. can be calculated via:
		\begin{equation}
		J= \int_{\text{l.o.s.}} ~\rho_\phi^2 (b,l,s) ds
		\end{equation}
		where $s$ is the distance away from the solar system, and $\rho_\phi$ is the DM density.

		DM not only exists in the Milky Way but also in other galaxies. The gamma-ray production on those galaxies will also contribute to the IGRB.
We consider extragalactic $\gamma$ rays to be isotropic. The flux received at Earth can be calculated via~\cite{cookbook}
		\begin{equation}
		\frac{dF_{\text{EG}\gamma}}{dE_\gamma}(E_\gamma)=
		\frac{c}{E_\gamma}\int_{0}^{\infty} dz \frac{1}{H(z)(1+z)}(\frac{1}{1+z})^3
		\frac{1}{8 \pi}
		B(z)(\frac{\bar{\rho}(z)}{m_\phi})^2
		\sum_l \langle\sigma v\rangle_l \frac{dN^l_\gamma}{dE_\gamma'}(E_\gamma')
		e^{-\tau(E_\gamma',z)}
		\end{equation}
		where $H(z)=H_0\sqrt{\Omega_m (1+z)^3+(1-\Omega_m)}$ is the Hubble function, $\bar{\rho}(z)=\bar{\rho}_0(1+z)^3$ is the average cosmological DM density, and $\bar{\rho}_0 \simeq 1.15\times 10^{-6}~\text{GeV}/\text{cm}^3$, $E_\gamma'=E_\gamma(1+z)$, $\tau(E_\gamma',z)$ is the optical depth, which is also provided numerically by PPPC 4 DM ID~\cite{cookbook} in the form of \textsc{Mathematica}\textsuperscript{\textregistered} interpolating functions. $\tau(E_\gamma',z)$ describes the absorption of gamma rays in the intergalactic medium between the redshifts $0$ and $z$. We calculated the Hubble function in the $\Lambda$CDM cosmology with a pressureless matter density of the Universe $\Omega_m=0.27$, dark energy density of the Universe $\Omega_\Lambda=0.73$, and a scale factor for the Hubble expansion rate $0.7$. $B(z)$ is a cosmological boost factor which takes the enhancement of the annihilation frequency by the DM cluster into account. In terms of the boost factor $B(z)$, we adopt the model supplied by Macci\`{o} et al.~\cite{boostBZ}. The results of the extragalactic gamma ray are shown in Fig.~\ref{FermiLATIGRBandJfactor} \texttt{(a)}; they show that the contribution of gamma rays from extragalactic DM can be neglected in high-energy regions.

	\paragraph{Constraining the position of the DM sub-halo from IGRB}

		Since the DM distribution in the Milky Way halo and DM subhalo is not isotropic for the Earth, the photon fluxes we receive in different l.o.s. are also different. Photon fluxes from different l.o.s. correspond to different J-factors. As can be seen from Fig.~\ref{FermiLATIGRBandJfactor} \texttt{(a)}, a prompt emission is dominant in the high-energy region; therefore, we use the condition $\breve{\chi}^2<4$ to set a one-sided $2\sigma$ C.L. limit on the J-factor. The $2\sigma$ C.L. upper bound of the J-factor is $4.9\times10^{22}~\text{GeV}^2/\text{cm}^5$. The size, distance, and density of the DM sub-halo would affect the J-factor. For the DM sub-halo that we used in Section~\ref{HintsfromDAMPE}, the J-factor of the l.o.s. which goes through the DM sub-halo center is shown in Fig.~\ref{FermiLATIGRBandJfactor} \texttt{(b)}. By checking Fig.~\ref{FermiLATIGRBandJfactor} \texttt{(b)}, we could conclude that the aperture angle between the l.o.s. going through the DM sub-halo center and the axis connecting the Earth to the galactic center should be $\theta > 41^\circ$.

	\paragraph{The tightest limit on $\langle\sigma v\rangle$ is from the galactic center}

		The DM density is large near the galactic center. Therefore, the diffuse $\gamma$-ray flux from the galactic center should set the tightest limits on the cross-section of DM. $3\sigma$ C.L. upper limits from Fermi-LAT $\gamma$-ray observations on the cross-section of the TeV leptophilic DM have been derived for several different windows on the sky~\cite{IGRBonWindow3}. Their results show that the window near the galactic center provides the tightest limit on the DM annihilation cross-section. In terms of the window $(0^\circ<|l|<8^\circ,1^\circ<|b|<9^\circ)$, the $3\sigma$ upper limits on $\langle\sigma v\rangle_{\phi,\phi\to \eta\to \mu^+,\mu^-}$ with $m_\phi=1.5~\text{TeV}$ is $8\times10^{-25}~\text{cm}^3/\text{s}$. Note that the cross-section required to fit the DAMPE data is only $3.6\times10^{-25}~\text{cm}^3/\text{s}$, which is smaller than the tightest limit.

	\subsection{Comparison with CALET}

		This section further compares our analysis with CALET data~\cite{CALET1}~\cite{CALET2}.
		The latest report from the CALET collaboration reported an electron and positron spectrum from 11 GeV to 4.8 TeV.

	\begin{figure}[htbp]
		\subfigure[]{\includegraphics[scale=0.43]{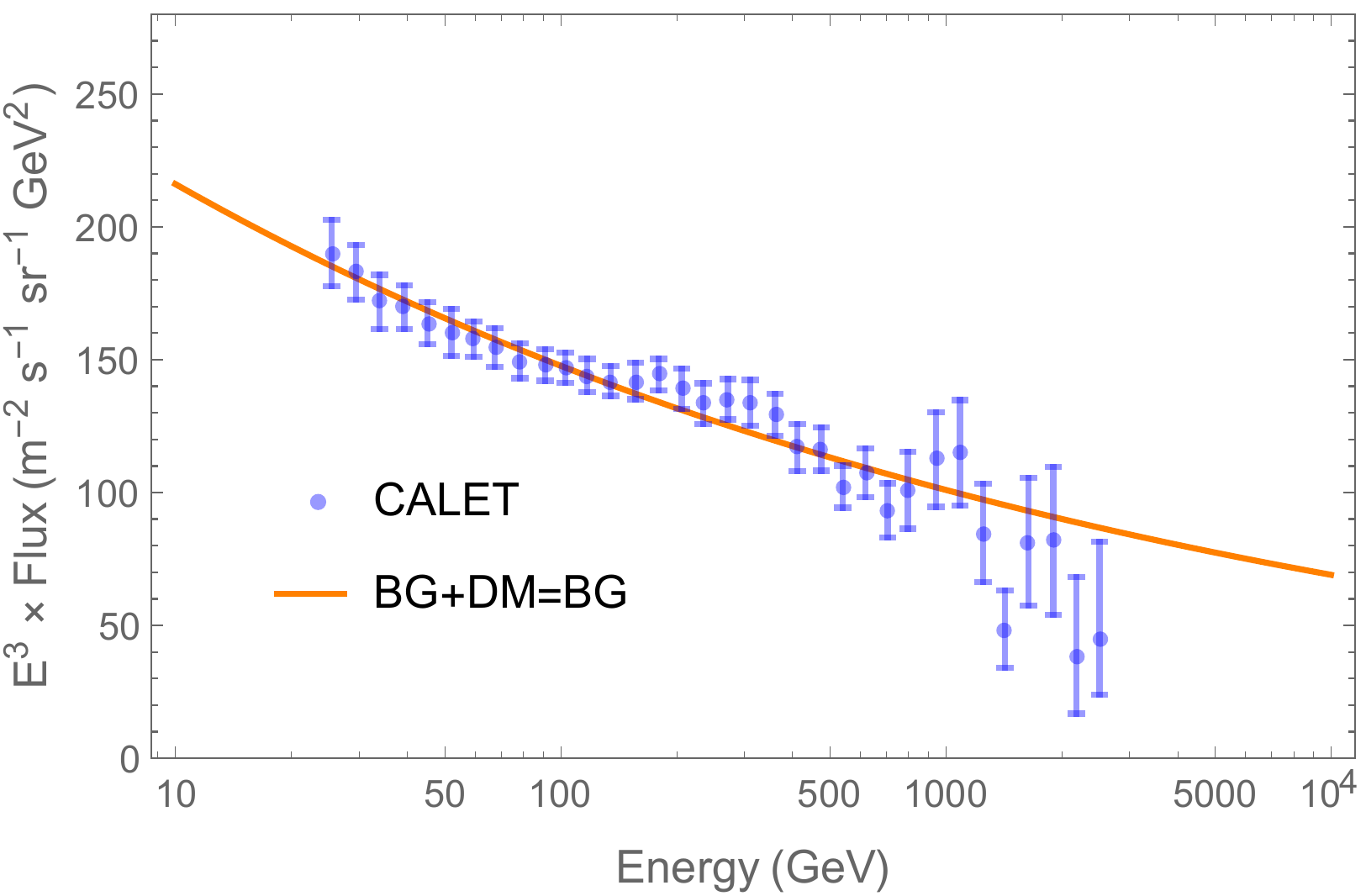}}
		\hfill
		\subfigure[]{\includegraphics[scale=0.43]{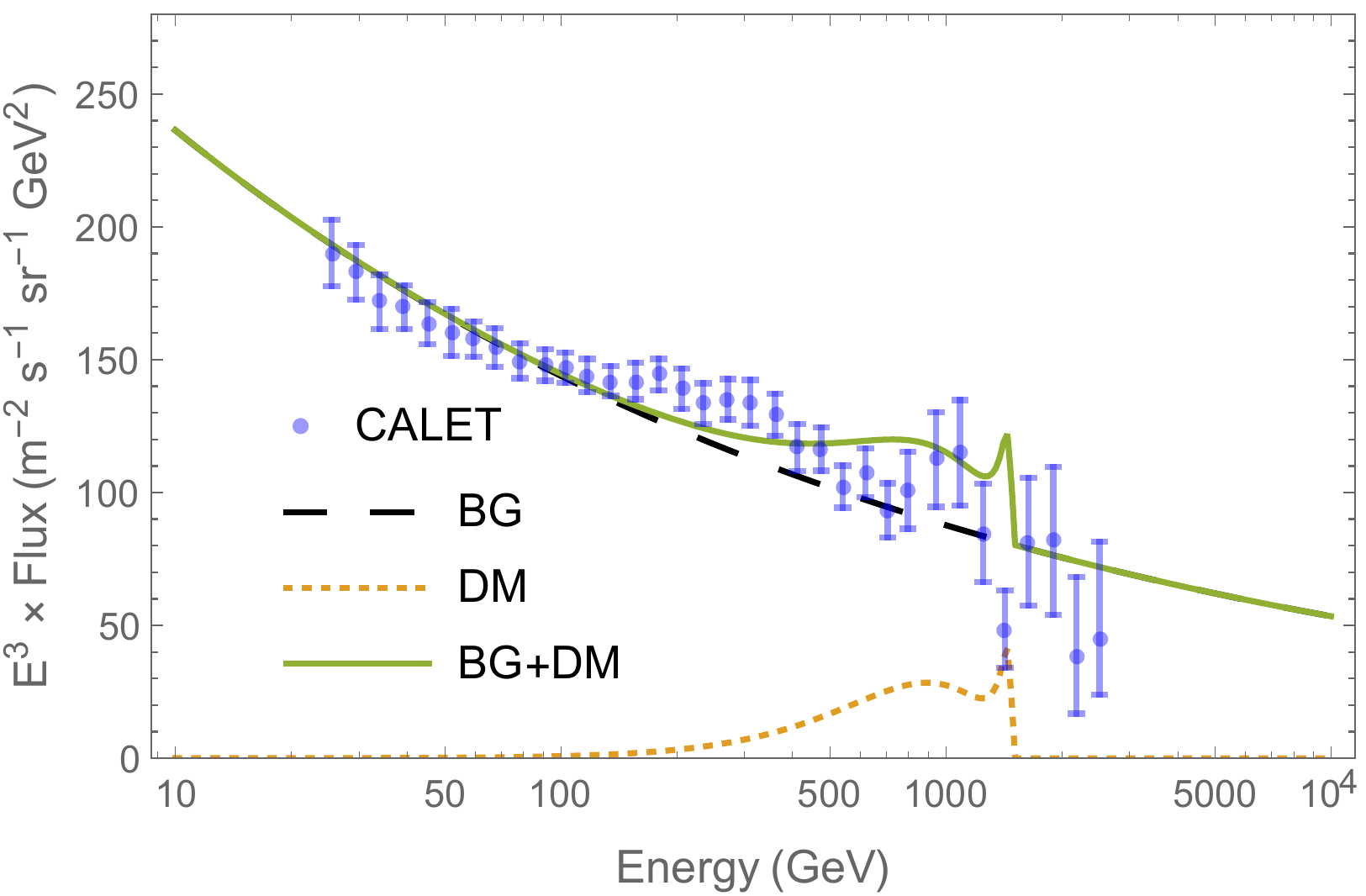}}
		\caption{\label{CALETfitandCALETfit2}
			Cosmic ray electron spectrum (multiplied by $E^3$). In both figures, the blue points with error bars are the flux measured by CALET.\\
			 \texttt{(a)} The combined case. The solid orange line shows the best-fitting spectrum. It does not support contributions from the DM. The fitting quality is $\chi^2/\text{d.o.f.}=33.6/35$.\\
			\texttt{(b)} The $\langle\sigma v\rangle$-fixed case. In the figure, it is presupposed that $m_\phi=1.5~\text{TeV}$, $\langle\sigma v\rangle_{\phi,\phi\to \eta\to e^+,e^-}=1.8\times10^{-26}~\text{cm}^3/\text{s}$, $\langle\sigma v\rangle_{\phi,\phi\to \eta\to \mu^+,\mu^-}=3.6\times10^{-25}~\text{cm}^3/\text{s}$. The dashed line shows the best-fitting cosmic electron background (BG). The fitting parameters are, $F^{\text{BG}}=C[ E/(100~\text{GeV})]^{-\alpha}$ with $C=1.44\times10^{-4}~(\text{GeV}~\text{m}^2~\text{s sr})^{-1}$ and $\alpha=3.22$, The fitting quality is $\chi^2/\text{d.o.f.}=67.2/37$. The dotted line shows the flux contributed by DM. The solid line represents the total flux.
		}
	\end{figure}

		For the CALET data, we also apply the $\chi^2$ fit to the following two cases:
		\begin{itemize}
			\item [\texttt{(a)}] The combined case. Lepton signals from DM annihilation are taken into account ($\langle\sigma v\rangle_{\phi,\phi\to\eta\to \bar{l},l}\geq0~\text{cm}^3/\text{s}$). The DM mass is fixed to $m_\phi=1.5~\text{TeV}$.
			\item [\texttt{(b)}] The $\langle\sigma v\rangle$-fixed case. It is preassumed that the contribution of DM is the same as that of DM in Fig.~\ref{DAMPEfitandDAMPEfit2} \texttt{(b)}; that is, the following parameters are taken: $m_\phi=1.5~\text{TeV}$, $\langle\sigma v\rangle_{\phi,\phi\to \eta\to e^+,e^-}=1.8\times10^{-26}~\text{cm}^3/\text{s}$, $\langle\sigma v\rangle_{\phi,\phi\to \eta\to \mu^+,\mu^-}=3.6\times10^{-25}~\text{cm}^3/\text{s}$.
		\end{itemize}
		With regard to the combined case, we perform a combined fit to all the points of CALET. The results reject the contribution from DM; namely, $\langle\sigma v\rangle_{\phi,\phi\to \eta\to e^+,e^-}=0~\text{cm}^3/\text{s}$, $\langle\sigma v\rangle_{\phi,\phi\to \eta\to \mu^+,\mu^-}=0~\text{cm}^3/\text{s}$. The results support a simple power-law spectrum, $F^{\text{BG}}=C[E/(100~\text{GeV})]^{-\alpha}$ with the best-fitting parameters: $C=1.48\times10^{-4}~(\text{GeV}~\text{m}^2~\text{s sr})^{-1}$ and $\alpha=3.17$. The fitting result is shown in Fig.~\ref{CALETfitandCALETfit2} \texttt{(a)}. The fitting quality is $\chi^2/\text{d.o.f.}=33.6/35$.

		Regarding the $\langle\sigma v\rangle$-fixed case, we find that contributions from DM particles with a mass of 1.5 TeV would make the fitting quality even worse. As illustrated by Fig.~\ref{CALETfitandCALETfit2} \texttt{(b)}, we adopt the same parameter for the DM sub-halo and the DM particles used in the fitting process of the DAMPE data and perform a fit to the CALET data. The result is $F^{\text{BG}}=C[E/(100~\text{GeV})]^{-\alpha}$, with the best-fitting parameters: $C=1.44\times10^{-4}~(\text{GeV}~\text{m}^2~\text{s sr})^{-1}$ and $\alpha=3.22$. The fitting quality is decreased from $\chi^2/\text{d.o.f.}=33.6/35$ (shown with Fig.~\ref{CALETfitandCALETfit2} \texttt{(a)}) to $\chi^2/\text{d.o.f.}=67.2/37$ (shown with Fig.~\ref{CALETfitandCALETfit2} \texttt{(b)}). This further reveals that the CALET data do not support the existence of the DM sub-halo.

		Although the data from DAMPE and CALET show different characteristics, when taking the large errors in high-energy lepton measurement data into consideration, further analysis should be performed when more data samples are collected.

	\section{Summary}

		This paper proposes a purely gravitational DM that can annihilate through the operators $-\xi_\eta (H^\dag \Phi + \Phi^\dag H) R$ and $-\xi_\phi \phi^2 R$. The newly introduced SU(2) doublet contains a scalar $\eta$, which is ultimately the key field for connecting the DM and leptons. The scalar mediator's mass $\eta$ is constrained to $m_\eta>m_Z/2$ by the kinetic term of $\Phi$. Since DM direct detection experiments aim to find DM particles or mediators below 1 GeV, those experiments impose almost no limits on our model. Then, we studied the $\eta$-on-shell process $\phi,\phi\to\eta, h$ and the $\eta$-off-shell process $\phi,\phi\to\eta\to \bar{l},l$.

		In terms of the $\eta$-on-shell process, it can be constrained by the observation of antiproton flux. The results show that the $\eta$-on-shell process is forbidden when the mass of DM particles is around the electroweak scale.

		Regarding the $\eta$-off-shell process, the $\langle\sigma v\rangle$ of the DM is sensitive to the DM temperature when $2m_\phi\approx m_\eta$. In this case, although the cross section changes with the evolution of the Universe, we can still obtain the correct relic density of the DM after introducing the operator $-\xi_h H^\dag HR$ and letting the cross-section of $\phi,\phi\to h,h$ channel take the appropriate value. This phenomenon also explains why the thermal average annihilation cross-section needed to obtain the correct excess in the cosmic lepton energy spectrum reported by DAMPE is many orders of magnitude larger than that needed to obtain the correct DM relic abundance.

		Unitarity bounds can also set constraints on the coupling constants of the model. The upper limits of $\xi_\eta$ and $\xi_\phi$ are $10^{14}\sim10^{15}$ for TeV DM. The cosmic positron flux, CMB anisotropy, and DM relic density can set constraints on $\xi_\phi\xi_\eta\xi_l$. The results show that there is a wide range of parameter space that is compatible with current experiments. The results also show that the DAMPE-favored region is not excluded.

	By assuming a nearby DM sub-halo with a DM mass of 1.5 TeV and choosing suitable parameters for the model, the predicted cosmic lepton energy spectrum can better fit the DAMPE data than a single power-law spectrum. In the better-fitted case, the broad excess in the cosmic lepton energy spectrum up to TeV energy is mainly contributed by $\phi,\phi\to\eta\to\mu^+,\mu^-$ channel. The line structure at about 1.4 TeV is mainly contributed by $\phi,\phi\to\eta\to e^+,e^-$ channel. The IGRB set strong constraints to the position of the DM sub-halo. The aperture angle between the direction of the sub-halo center and the axis connecting the Earth to the galactic center depends on the size, distance, and density of the sub-halo; in our scenario, it should be at least $41^\circ$ at $2\sigma$ C.L.. We also find that the CALET data do not support the existence of a DM sub-halo. Therefore, further analysis should be undertaken after DAMPE and CALET have collected enough samples.

	\acknowledgments

	Supported by the National Key Research and Development Program 2018YFA0404204, the National Science Foundation of China (U1531131 and U1738211), and the Science Foundation of Yunnan Province (NO. 2018FA004). We also thank the reviewer for doing a lot of research and contributing about half of the references for this work.

\end{document}